\RequirePackage{amsmath}
\documentclass[12pt,a4paper]{iopart}

\usepackage{graphicx}
\usepackage{cite}
\usepackage{color}
\usepackage{multirow}
\usepackage{epsfig}
\usepackage{epstopdf}
\usepackage{iopams}
\usepackage{cancel}
\usepackage{multicol}
\usepackage{ipa}
\usepackage{mathabx}
\usepackage{setspace}

\usepackage[breaklinks=true,colorlinks=true,linkcolor=blue,urlcolor=blue,citecolor=blue]{hyperref}

\usepackage[deletedmarkup=xout]{changes}
\definechangesauthor[color=red]{RPB}
\definechangesauthor[color=orange]{SW}
\definechangesauthor[color=green]{ZD}
\definechangesauthor[color=magenta]{MK}

\begin{document}

\title[Non-neutral discharge regime of RF plasma jets]{A non-neutral regime of radio-frequency atmospheric pressure  plasma jets: \\ Simulation and modeling}

\author{Maximilian Klich$^1$, Sebastian Wilczek$^1$, Zolt\'{a}n Donk\'{o}$^2$, Ralf Peter Brinkmann$^1$}

\address{$^1$Ruhr University Bochum, Department of Electrical Engineering and Information Science, 44780 Bochum, Germany}
\address{$^2$ Institute for Solid State Physics and Optics, Wigner Research Centre for Physics, Konkoly-Thege Miklós str. 29-33, H-1121 Budapest, Hungary}

\date{\today}

\vspace{4mm}

\noindent M.~Klich: \href{https://orcid.org/0000-0002-3913-1783}{https://orcid.org/0000-0002-3913-1783}

\noindent S.~Wilczek: \href{https://orcid.org/0000-0003-0583-4613}{https://orcid.org/0000-0003-0583-4613}

\noindent Z.~Donk\'{o}: \href{https://orcid.org/0000-0003-1369-6150}{https://orcid.org/0000-0003-1369-6150}

\noindent R.P.~Brinkmann: \href{https://orcid.org/0000-0002-2581-9894}{https://orcid.org/0000-0002-2581-9894}

\begin{abstract}
    Radio-frequency-driven atmospheric pressure plasma jets (RF APPJs) play an essential role in many technological applications.
    This work studies the characteristics of these discharges in the so-called non-neutral regime where the conventional structure of a quasi-neutral bulk and an electron depleted sheath does not develop, and the  electrons are instead organized in a drift-soliton-like structure that never reaches quasi-neutrality. 
	A  hybrid particle-in-cell/Monte Carlo collisions (PIC/MCC) simulation is set up, which combines a fully kinetic electron model via the PIC/MCC algorithm with a drift-diffusion model for the ions. 
    In addition, an analytical model for the electron dynamics is formulated.
	The formation of the soliton-like structure and the connection between the soliton and the electron dynamics are investigated.
	The location of the electron group follows a drift equation, while the spatial shape can be described by  Poisson-Boltzmann equilibrium in a co-moving frame.
	A stability analysis is conducted using the Lyapunov method and a linear stability analysis.
	A comparison of the numerical simulation with the analytical models yields a good agreement.
\end{abstract}


\pagebreak

\section{Introduction}
	
	Non-equilibrium plasmas have become a versatile and commonly used tool in modern society \cite{LiebermanLichtenberg2005, MakabePetrovic2015, Becker2005, Fridman2012, Raizer1991}.
	The origins of gas discharges trance back to discharges  ignited under atmospheric pressure conditions.
	These first human-created plasmas root back to the 19th century \cite{Raizer1991, Bruggeman2017, Kogelschatz2003}.
	Recently, those discharges were re-established as the focus of many works and considered to be the solution to fundamental challenges of the 21st century.
	The proposed applications range from environmental and energy-related topics such as CO\textsubscript{2} conversion \cite{Bogaerts2015, George2021} to medical applications such as cancer treatment \cite{Keidar2015, Yan2017} or wound healing \cite{Xu2015, Bekeschus2016}.
	(A far more complete list of applications is given by Adamovich et al. \cite{Adamovich2017}).
	The foundation of all applications of atmospheric pressure plasmas is their characteristic to enable complex chemistry without thermal reactions \cite{LiebermanLichtenberg2005, Fridman2012, Bruggeman2017, Kogelschatz2003, Adamovich2017}.
	For the biological surfaces in medical applications \cite{Keidar2015, Yan2017, Xu2015, Bekeschus2016} and other heat-sensitive materials \cite{LiebermanLichtenberg2005, MakabePetrovic2015, Becker2005, Bruggeman2017}, this characteristic of non-equilibrium plasmas allows for treatment at all.\par
	While the above-mentioned reasoning is proper for both atmospheric pressure and low-pressure plasmas, the most significant difference between both types of discharges is the vacuum chamber.
	The opportunity to be installed in ambient air renders an atmospheric pressure plasma application, in most cases, cheaper and the technical requirements less sophisticated \cite{Bruggeman2017}.
	Additionally, applications that involve biological probes, bodily tissues, or even a whole patient \cite{Bruggeman2017, Keidar2015, Yan2017, Xu2015, Bekeschus2016, Adamovich2017} cannot be realized under conditions anyhow nearby to vacuum.
	Nevertheless, atmospheric pressure non-equilibrium plasmas have specific difficulties too.
	Most notable in this context is the prevalence to transit, if provided with sufficient power, to a local thermodynamic equilibrium \cite{Becker2005, Fridman2012, Bruggeman2017}.
	Numerous different concepts for plasma sources and even more realizations have been utilized to circumvent this difficulty \cite{Becker2005, Bruggeman2017, Golda2016}.\par
	Standardized plasma sources are called for to limit the vast parameter space.
	Therefore, the European COST (Cooperation in Science and Technology) Action MP1011 on 'Biomedical Applications of Atmospheric Pressure Plasma Technology' provided Golda et al. \cite{Golda2016} with the opportunity to build the COST Reference Microplasma Jet (COST-Jet) as a reference source \cite{COST}.
	In the following, the COST-Jet has proven to be highly reproducible \cite{Riedel2020}, suitable for both experimental and theoretical studies \cite{Gorbanev2018}, and surpassed the status of "just a reference" source \cite{Gorbanev2019}.
	Hence, we choose the geometry of the COST-Jet for our models and simulations.\par
	Despite many advances and developments in recent years \cite{Bruggeman2017, Adamovich2017}, some basic principles, reaction pathways, in short, a complete and fundamental understanding of atmospheric pressure non-equilibrium plasmas remains desirable.
	The intrinsic complexity of the non-linear system of non-equilibrium plasmas is coupled to transient physics and inherently complex chemistry at ambient pressure.
	As a result of this contraction, theory and simulations struggle to keep up with the challenges.
	To completely describe the tightly bounded capacitively coupled microplasmas present at atmospheric pressures, a multi-time-scale, kinetic, and three-dimensional model that considers all chemical details is needed.
	Nevertheless, all these at once are currently simply not feasible.
	A complete description of the chemistry involves dozens of species and hundreds of reactions \cite{Becker2005, Fridman2012, Kemaneci2016, Koelman2017, Waskoenig2010, Niemi2011}.
	Models trying to resolve as many chemical details as possible are either global models with no spatial resolution \cite{Kemaneci2016, Koelman2017} or at maximum one-dimensional fluid models \cite{Waskoenig2010, Niemi2011}.
	Models prioritizing the spatial resolution result in two-dimensional fluid simulations \cite{Shaper2009, Hemke2011, Kelly2014} or one-dimensional kinetic models \cite{Eremin2016, Donko2018} and restrict their chemistry to the essential minimum. \par
	As the topical review by Bruggeman et al. \cite{Bruggeman2017} stresses, it is common knowledge that atmospheric pressure discharges do not necessarily form a quasi-neutral plasma bulk.
	This effect stems from the scaling of the Debye length $\lambda_\mathrm{D}$ compared to the discharge length $L$.
	Boundary effects (e.g., sheath formation) take place on the length scale of $\lambda_\mathrm{D} \approx 10^{-1} - 10^{-2}\,$mm and disturb the quasi-neutrality the bulk plasma strives for.
	For low-pressure plasmas, $L$ is in the range of several centimeters and boundary effects govern just a small part of the plasma.
	However, atmospheric pressure plasmas use small discharge lengths $L$ to avoid thermalization.
	Boundary effects may consequently affect the whole discharge and prevent the formation of a quasi-neutral bulk region.
	We chose to call a discharge regime where no quasi-neutral plasma bulk is formed a non-neutral discharge regime.\par
	Although we suspect to see it within a He/O\textsubscript{2} mixture in previous work \cite{Hemke2011}, a non-neutral discharge regime of the COST-Jet has not been analyzed in details.
	Employing a one-dimensional hybrid particle-in-cell/Monte Carlo collisions (PIC/MCC) simulation, we explore a non-neutral discharge for the He/N\textsubscript{2} mixture.
	Based on the simulation data, this work aims to describe a non-neutral discharge regime for the COST-Jet.
	An emphasis is made on the electron dynamics of this regime.
	We find the electrons organized within a moving Gaussian-shaped spatial profile, which features an immense form-stability.\par
	We, according to its tendency to keep and quickly regain its shape, characterize this structure as drift-soliton-like.
	The expression soliton describes wave phenomena with a high degree of dimensional stability that occur in many fields of physics  \cite{Drazin1989, Rajaraman1982}.
	In the context of plasma applications, actual drift-solitons have mainly been reported for magnetized plasmas \cite{Zhmudsky1970, Lakhin1988}.
	We find the plasma properties such as the electron temperature $T_\mathrm{e}$ dominated by the dynamics of the soliton-like structure.
	However, the hybrid PIC/MCC simulation functions in this regard more like a numerical experiment and is too sophisticated to develop a fundamental of the formed soliton.
	Therefore, an abstract and simplified analytical model for the electron dynamics is formulated.
	A detailed analysis of this model helps to explain the formation and spatial mold of the electron group and provides insights to its dominant influence on the electron dynamics.
	Additionally, deliberations on the stability of the analytical solutions that include the ideas of Lyapunov \cite{Lyapunov1966, Lyapunov1992, Pukdeboon2011} will be used to argue for the soliton-analogy. \par

	The rest of the manuscript is organized as follows.
	Section \ref{COSTjet} gives a brief overview of the COST-Jet and describes the chemistry
	under investigation.
	The following section introduces our simulation framework and essential details of its realization.
	The simulation-data-based analysis of the electron dynamics follows in section \ref{simulation}.
	Section \ref{modeling} will provide the aforementioned analytical model that provides generalization and further insight.
	The analysis of the model is acompanied by deliberations on the stability of the soliton solution.
	A comparison between the simulation and our analytical model is made in section \ref{comparison} and aims to visualize the reliability of the analytical model.
	The final section summarizes our findings, draws conclusions, and outlines perspectives for future work.
	
\clearpage
\newpage


\section{The COST reference Jet} \label{COSTjet}

\begin{figure}[t!h!]
		\centering
		\includegraphics[width = \textwidth]{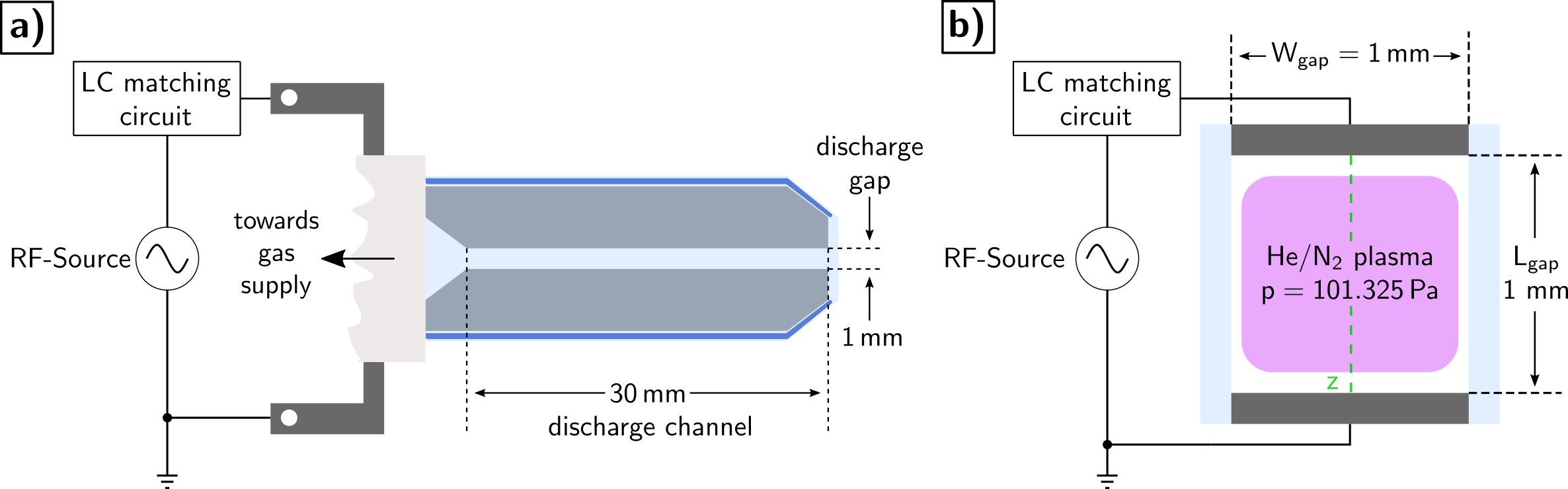}
		\caption{Simplified schematic sketches of two cross-sections of the COST-Jet (refer to Golda et al. \cite{Golda2016} for complete schematics and description).
		Electrodes are depicted in dark gray, and the quartz dielectric is shown in light blue. a) A cross-section in the $xz$-plane along the gas flow direction $x$.
		b) A cross-section in the $yz$-plane (perpendicular to the gas flow direction) at an arbitrary position of the discharge channel.
		The dashed green line marks the position of the $z$-axis in the center of the discharge along which the simulation is run.}
		\label{fig:jet}
	\end{figure}

	The COST-Jet is a capacitively coupled radio-frequency-driven plasma jet specifically designed to function as a reference plasma source.
	Nevertheless, recent work \cite{Gorbanev2019} shows that the COST-Jet is a more than decent plasma source for many applications.
	While Golda et al. \cite{Golda2016} provide a complete characterization and definition of the COST-Jet, this section gives a brief introduction to the device's essential features.
	By design, the COST-Jet is optimized for diagnostic purposes, especially optical access and is suitable for theoretical description.
	This choice is reflected in the two quartz plates glued to two stainless steel electrodes to form the rectangular discharge channel.
	The quartz plates allow for optical diagnostics along the horizontal gas-flow axis $x$ and vertical axis $z$.
	Figure \ref{fig:jet} a) shows a sketch of the cross-section of the COST-Jet along the gas-flow-axis $x$.
	In addition, the capacitively coupled power delivery at 13.56$\,$MHz combined with the rectangular geometry allows one-dimensional simulations to give meaningful results \cite{Shaper2009, Waskoenig2010, Niemi2011, Korolov2019, Korolov2020}.\par

	Although we firmly believe that a complete model of the Cost-Jet, a long tube ($30\, \mathrm{mm}$) with a tightly bounded $1\,\mathrm{mm} \times 1\, \mathrm{mm}$ lumen, has to regard all three spatial dimensions somehow, technical limitations currently force us to focus our study on a one-dimensional model.
	Similar to previous studies \cite{Shaper2009, Waskoenig2010, Niemi2011, Korolov2019, Korolov2020}, we focus on describing the plasma between the electrodes along the $z$-axis and perpendicular to the gas flow.
	Figure \ref{fig:jet} b) shows a cross-section of the setup, and the green line marks the position of the $z$-axis.
	For now, the gas flow is neglected and the plasma is assumed to be invariant in both the $x$- and $y$- direction.\par

	\begin{table}[t!]
		\centering
		\begin{tabular*}{1.0\textwidth}{ l @{\extracolsep{\fill}} l l c c}
			\hline
			No.~ & reaction & process name & $\varepsilon_\mathrm{thr}$ / $k_\mathrm{r}$ &  src. \\
			\hline
			1 & $\mathrm{e} + \mathrm{He} \rightarrow \mathrm{e} + \mathrm{He}$ & elastic scattering & - & (1) \\
			2 & $\mathrm{e} + \mathrm{He} \rightarrow \mathrm{e} + \mathrm{He}^\ast$ & triplet excitation & $19.82\, \mathrm{eV}$ & (1)\\
			3 & $\mathrm{e} + \mathrm{He} \rightarrow \mathrm{e} + \mathrm{He}^\ast$ & singlet excitation & $20.61\, \mathrm{eV}$ & (1)\\
			4 & $\mathrm{e} + \mathrm{He} \rightarrow \mathrm{e} + \mathrm{He}^+ + \mathrm{e}$ & ionization & $24.59\, \mathrm{eV}$ & (1)\\
			\hline
			5 & $\mathrm{e} + \mathrm{N}_2 \rightarrow \mathrm{e} + \mathrm{N}_2$ & elastic scattering & - & (2)\\
			6 & $\mathrm{e} + \mathrm{N}_2 \rightarrow \mathrm{e} + \mathrm{N}_2\,(r=\ast)$ & rot. excitation & $20\, \mathrm{meV}$ & (2)\\
			7 & $\mathrm{e} + \mathrm{N}_2 \rightarrow \mathrm{e} + \mathrm{N}_2\,(v=1)$ & vib. excitation & $290\, \mathrm{meV}$ & (2)\\
			8 & $\mathrm{e} + \mathrm{N}_2 \rightarrow \mathrm{e} + \mathrm{N}_2\,(v=1)$ & vib. excitation & $291\, \mathrm{meV}$ & (2)\\
			9 & $\mathrm{e} + \mathrm{N}_2 \rightarrow \mathrm{e} + \mathrm{N}_2\,(v=2)$ & vib. excitation & $590\, \mathrm{meV}$ & (2)\\
			10 & $\mathrm{e} + \mathrm{N}_2 \rightarrow \mathrm{e} + \mathrm{N}_2\,(v=3)$ & vib. excitation & $880\, \mathrm{meV}$ & (2)\\
			11 & $\mathrm{e} + \mathrm{N}_2 \rightarrow \mathrm{e} + \mathrm{N}_2\,(v=4)$ & vib. excitation & $1.17\, \mathrm{eV}$ & (2)\\
			12 & $\mathrm{e} + \mathrm{N}_2 \rightarrow \mathrm{e} + \mathrm{N}_2\,(v=5)$ & vib. excitation & $1.47\, \mathrm{eV}$ & (2)\\
			13 & $\mathrm{e} + \mathrm{N}_2 \rightarrow \mathrm{e} + \mathrm{N}_2\,(v=6)$ & vib. excitation & $1.76\, \mathrm{eV}$ & (2)\\
			14 & $\mathrm{e} + \mathrm{N}_2 \rightarrow \mathrm{e} + \mathrm{N}_2\,(v=7)$ & vib. excitation & $2.06\, \mathrm{eV}$ & (2)\\
			15 & $\mathrm{e} + \mathrm{N}_2 \rightarrow \mathrm{e} + \mathrm{N}_2\,(v=8)$ & vib. excitation & $2.35\, \mathrm{eV}$ & (2)\\
			16 & $\mathrm{e} + \mathrm{N}_2 \rightarrow \mathrm{e} + \mathrm{N}_2\,(A\, ^3\Sigma_\mathrm{u}^+, v=\text{0$\,$-$\,$4})$ & electr. excitation & $6.17\, \mathrm{eV}$ & (2)\\
			17 & $\mathrm{e} + \mathrm{N}_2 \rightarrow \mathrm{e} + \mathrm{N}_2\,(A\, ^3\Sigma_\mathrm{u}^+, v=\text{5$\,$-$\,$9})$ & electr. excitation & $7.00\, \mathrm{eV}$ & (2)\\
			18 & $\mathrm{e} + \mathrm{N}_2 \rightarrow \mathrm{e} + \mathrm{N}_2\,(B\, ^3\Pi_\mathrm{g})$ & electr. excitation & $7.35\, \mathrm{eV}$ & (2)\\
			19 & $\mathrm{e} + \mathrm{N}_2 \rightarrow \mathrm{e} + \mathrm{N}_2\,(W\, ^3\Delta_\mathrm{u})$ & electr. excitation & $7.36\, \mathrm{eV}$ & (2)\\
			20 & $\mathrm{e} + \mathrm{N}_2 \rightarrow \mathrm{e} + \mathrm{N}_2\,(A\, ^3\Sigma_\mathrm{u}^+, v \geq 10)$ & electr. excitation & $7.80\, \mathrm{eV}$ & (2)\\
			21 & $\mathrm{e} + \mathrm{N}_2 \rightarrow \mathrm{e} + \mathrm{N}_2\,(B^\prime \, ^3\Sigma_\mathrm{u}^-)$ & electr. excitation & $8.16\, \mathrm{eV}$ & (2)\\
			22 & $\mathrm{e} + \mathrm{N}_2 \rightarrow \mathrm{e} + \mathrm{N}_2\,(a^\prime \, ^1\Sigma_\mathrm{u}^-)$ & electr. excitation & $8.40\, \mathrm{eV}$ & (2)\\
			23 & $\mathrm{e} + \mathrm{N}_2 \rightarrow \mathrm{e} + \mathrm{N}_2\,(a\, ^1\Pi_\mathrm{g})$ & electr. excitation & $8.55\, \mathrm{eV}$ & (2)\\
			24 & $\mathrm{e} + \mathrm{N}_2 \rightarrow \mathrm{e} + \mathrm{N}_2\,(w\, ^1\Delta_\mathrm{u})$ & electr. excitation & $8.89\, \mathrm{eV}$ & (2)\\
			25 & $\mathrm{e} + \mathrm{N}_2 \rightarrow \mathrm{e} + \mathrm{N}_2\,(C\, ^3\Pi_\mathrm{u})$ & electr. excitation & $11.03\, \mathrm{eV}$ & (2)\\
			26 & $\mathrm{e} + \mathrm{N}_2 \rightarrow \mathrm{e} + \mathrm{N}_2\,(E\, ^3\Sigma_\mathrm{g}^+)$ & electr. excitation & $11.87\, \mathrm{eV}$ & (2)\\
			27 & $\mathrm{e} + \mathrm{N}_2 \rightarrow \mathrm{e} + \mathrm{N}_2\,(a^{\prime\prime}\, ^1\Sigma_\mathrm{g}^+)$ & electr. excitation & $12.25\, \mathrm{eV}$ & (2)\\
			28 & $\mathrm{e} + \mathrm{N}_2 \rightarrow \mathrm{e} + \mathrm{N} + \mathrm{N}$ & dissociation & $13.0\, \mathrm{eV}$ & (2)\\
			29 & $\mathrm{e} + \mathrm{N}_2 \rightarrow \mathrm{e} + \mathrm{N\textsubscript{2}\textsuperscript{+}}$ & ionization & $15.6\, \mathrm{eV}$ & (2)\\
			30 & $\mathrm{e} + \mathrm{N}_2 \rightarrow \mathrm{e} + \mathrm{N\textsubscript{2}\textsuperscript{+}}$ & ionization & $18.8\, \mathrm{eV}$ & (2)\\
			\hline
			31 & $\mathrm{He}^\ast + \mathrm{N}_2 \rightarrow \mathrm{He} + \mathrm{N\textsubscript{2}\textsuperscript{+}} + \mathrm{e}$ & Penning ionization & $5.0 \times 10^{-17}\,$m\textsuperscript{3}/s & (3)\\
			32 & $\mathrm{He}^+ + \mathrm{He} + \mathrm{He} \rightarrow \mathrm{He\textsubscript{2}\textsuperscript{+}} + \mathrm{He}$ & ion conversion & $1.1 \times 10^{-43}\,$m\textsuperscript{6}/s & (3)
		\end{tabular*}
		\caption{Plasma chemical reactions considered in the simulation. The last column uses the following abbreviations for the data sources of the corresponding cross sections or reaction rates: (1) data from the \textit{Biagi-v7.1} database obtained via the website of the LXCat project \cite{lxcat1, lxcat2, lxcat3}, (2) data based on the measurements of Phelps and Pitchford \cite{Phelps1985} also obtained via the website of the LXCat project \cite{lxcat1, lxcat2, lxcat3}, (3) reaction rates retrieved from Brok et al. \cite{Brok2005} and Sakiyama and Graves \cite{Sakiyama2006}. The second column from the right gives threshold energies $\varepsilon_\mathrm{thr}$ for electron reactions and reaction rates $k_\mathrm{r}$ for ion reactions.}
		\label{tab:chemistry}
	\end{table}
	
	In terms of chemistry, the COST-Jet is operated using a chemically inert carrier gas (usually helium \cite{Golda2016, Shaper2009, Waskoenig2010, Niemi2011, Hemke2011, Kelly2014, Korolov2019, Korolov2020}, less commonly argon \cite{Gorbanev2018, Golda2020}) mixed with a molecular trace gas (usually nitrogen \cite{Shaper2009, Korolov2019, Korolov2020, Golda2020} and/or oxygen \cite{Golda2016, Waskoenig2010, Niemi2011, Hemke2011, Kelly2014, Gorbanev2018, Gorbanev2019, Golda2020, Mouchtouris2021}).
	For this study, we decided to examine the helium/nitrogen mixture.
	The resulting electropositive plasma yields a moderate amount of species and chemical complexity.
	Both are beneficial for the analytical model that will be presented in a following section.
	The details of the chemistry set used for this study are mainly oriented on previous work \cite{Donko2018} and are shown in table \ref{tab:chemistry}.

\newpage


\section{Hybrid particle-in-cell/Monte Carlo collisions simulation} \label{eehric}
	
	The particle-in-cell (PIC) simulation is a numerically sophisticated tool that provides self-consistently calculated statistical representations of the distribution functions \cite{HockneyBook, Birdsall, Verboncoeur2005, Wilczek2020}.
	The foundation of the PIC algorithm traces back to the 1940s \cite{HockneyBook}.
	In the 1960s, a Monte Carlo collisions (MCC) scheme was added that lead to the commonly used particle-in-cell/Monte Carlo collisions (PIC/MCC) simulation \cite{Birdsall}.
	In recent years, the PIC/MCC simulation has had great success in the research of non-equilibrium plasmas in general \cite{LiebermanLichtenberg2005, VanDijk2009, Verboncoeur2005, Wilczek2020, PICbenchmark}, and the scheme has been applied in the atmospheric pressure regime \cite{Korolov2019, Korolov2020, Donko2018}.
	Nevertheless, the above-discussed demands of atmospheric pressure discharges severely drive PIC/MCC simulations to their limits.
	Therefore, hybrid PIC/MCC simulation schemes for RF-driven plasma jets are utilized to lower the computational burden while keeping all essential pieces of physics \cite{Eremin2016, Liu2020, VanDijk2009}.\par
	Overall, hybrid concepts in plasma modeling are commonly used.
	The review paper of van Dijk et al. \cite{VanDijk2009} lists numerous techniques, variations, and applications.
	The hybrid PIC simulation has many applications in astrophysical plasma studies \cite{Harned1982, Hara2012, Nunn1993, Kubota2016}.
	These studies reduce the numerical load of their calculations by describing the electrons as a fluid and solely resolving the ion dynamics kinetically.
	As Eremin et al. \cite{Eremin2016} point out, the situation in RF plasma jets is precisely inverted.
	Electrons bear dynamics that require a fully kinetic description, while a fluid description of the ion dynamics leads to a significantly more efficient model.\par
	For this study, we adapt the hybrid PIC/MCC simulation concept and develop the one-dimensional \underline{e}xplicit \underline{e}lectrostatic \underline{H}ybrid pa\underline{r}ticle-\underline{i}n-\underline{c}ell/Monte Carlo collisions code \textit{eehric}.
	The following subsections present some details of the implementation.
	
	\subsection{Description of the electrons} \label{electrons}
		
		As in the classical PIC/MCC scheme, the electrons are described as particles, represented by so-called superparticles.
		Their dynamics are traced by solving the equations of motion individually for each superparticle \cite{HockneyBook, Birdsall, Verboncoeur2005, Wilczek2020}.
		The code \textit{eehric} applies the Leap-Frog algorithm on a regular one-dimensional Cartesian discretization of the equations of motion.
		The width $\Delta{z}$ of the resulting spatial grid is coupled to the Debye length by $\lambda_\mathrm{D} \gtrsim 2\, \Delta{z}$ to avoid numerical instabilities \cite{Birdsall, Wilczek2020, PICbenchmark}.
		For the Cost-Jet geometry, the small electrode separation allows a low number of grid points to satisfy the requirement.
		We fixed the number of grid points for this study to 201.
		The usual Monte Carlo technique is implemented to account for collisions \cite{Birdsall, Wilczek2020}.
		Additionally, a null-collision technique \cite{Skullerud} enhances the numerical efficiency of the collision routine.
		The necessary cross sections for the electron processes listed in table \ref{tab:chemistry} are retrieved from the website of the LXCat project \cite{lxcat1, lxcat2, lxcat3}.
		For helium, the cross section data originate from the  database \textit{Biagi-v7.1} based on version 7.1 of the simulation program \textit{Magboltz} \cite{Biagi1989}.
		For the nitrogen cross sections, the Phelps database provides data based on measurements by Phelps and Pitchford \cite{Phelps1985}.
		In terms of this study, as well as in previous work \cite{Donko2018}, the effects of gas heating are neglected.\par
		The collision probability for electrons $P_\mathrm{e}$ is found by 
		\begin{align}
			P_\mathrm{e} = 1 - \exp{(- \nu_\mathrm{e}\, \Delta{t})},
		\end{align}
		where $\nu_\mathrm{e}$ is the electron collision frequency and $\Delta{t}$ the discrete timestep \cite{Birdsall, Wilczek2020, PICbenchmark, Donko2018}.
		The criterion for stability $\Delta{t}\, \omega_\mathrm{pe} \lesssim 0.2$ that requires the time step $\Delta{t}$ to sufficiently resolve the electron plasma frequency $\omega_\mathrm{pe}$ is overshadowed by the criterion $\nu_\mathrm{e}\, \Delta{t} \ll 1$ \cite{Birdsall, Wilczek2020, PICbenchmark}.
		The latter criterion follows from a requirement that forbids particles to suffer consecutive collisions per time step.
		Under atmospheric pressure conditions, this criterion enforces a much smaller time step than the stability requirement.
		We use 1.2 million time steps for each RF cycle to limit the probability for a second (i.e., missed) collision per time step to approximately 1 percent.
		The results shown in later sections are averaged over 2000 RF periods and binned to 1000 diagnostic time steps within an RF period.
		This averaging process allows for satisfactory results with a relatively low number of about 10000 superparticles.
		As Erden and Rafatov \cite{Erden2014} present, the influence of the number of superparticles on the statistics always has to be considered.

	\subsection{Description of the ions} \label{ions}
		
		Eremin et al. present that a drift-diffusion approximation is a valid representation of the ion dynamics at high pressures \cite{Eremin2016}.
		Thus, \textit{eehric} adapts this insight and simulates the He\textsuperscript{+}, He\textsubscript{2}\textsuperscript{+}, and N\textsubscript{2}\textsuperscript{+} ions by evaluating the following form of the one-dimensional drift-diffusion approximation:
		\begin{align}
			\frac{\partial{n_{\mathrm{i,}s}}}{\partial{t}} + \frac{\partial{\Gamma_{\mathrm{i,}s}}}{\partial{z}} &= S_{\mathrm{i,}s}, \label{eq:DD1}\\
			\Gamma_{\mathrm{i,}s} &= n_{\mathrm{i,}s}\, \mu_{\mathrm{i,}s}\, E_z - D_{\mathrm{i,}s}\, \frac{\partial{n_{\mathrm{i,}s}}}{\partial{z}}. \label{eq:DD2}
		\end{align}
		The index $s$ represents the individual ion species.
		$n_\mathrm{i}$ is the ion number density, $\Gamma_\mathrm{i}$ is the ion flux density, $\mu_\mathrm{i}$ is the mobility constant, $E_z$ denotes the electric field, and $D_\mathrm{i}$ is the mobility constant. \par
		As argued in previous work \cite{Donko2018}, the trace gas admixture is usually one percent or lower.
		Thus, the influence of nitrogen on the gas transport is negligible, and the binary diffusion coefficients  and ion mobility in helium yield a sufficiently accurate description.
		The necessary mobility data have been measured by Frost \cite{Frost1957} (He\textsuperscript{+}/He), Beaty and Patterson \cite{Beaty1965} (He\textsubscript{2}\textsuperscript{+}/He), and McFarland et al. \cite{McFarland1973} (N\textsubscript{2}\textsuperscript{+}/He).
		The numerical values for He\textsubscript{2}\textsuperscript{+}/He and N\textsubscript{2}\textsuperscript{+}/He are retrieved from the data collection of Ellis et al. \cite{Ellis1976}.
		Values for the binary diffusion coefficients have been obtained by Deloche et al. \cite{Deloche1976} (He\textsuperscript{+}/He and He\textsubscript{2}\textsuperscript{+}/He) and Walker and Westenberg \cite{Walker1958} (N\textsubscript{2}\textsuperscript{+}/He).\par
		For the numerical solution of equations \eqref{eq:DD1} and \eqref{eq:DD2}, the scheme presented by Scharfetter and Gummel \cite{Scharfetter1969} is applied.
		Other authors \cite{Patankar1980} refer to this scheme as exponential scheme due to its exponential nature that provides an inherent switching between upwind and downwind differencing. \par
		Ion-induced secondary electron emission is included by basically counting the number of secondary electrons $N_\mathrm{SE,r/l}$ at each electrode separately and releasing electrons as long as $N_\mathrm{SE,r/l} \geq 1$.
		To account for an in time average sufficient number of secondary electrons, the decimal part of $N_\mathrm{SE,r/l}$ are treated by a random process.
		Whenever the equally distributed pseudo-random number in the interval $\left[ 0,1 \right)$ $R_{01} \leq N_\mathrm{SE,r/l}$, a secondary electron is generated at the respective electrode (left/right).
		The total number of secondary electrons is calculated by
		\begin{align}
			N_\mathrm{SE,r/l} =  \frac{1}{w_\mathrm{e}}\, \sum_s \gamma_{\mathrm{i},s}\, \Gamma_{\mathrm{i,el},s}\, \frac{\Delta{t}}{\Delta{z}}.
		\end{align}
		$w_\mathrm{e}$ is the weight of the electron superparticles, $\gamma_{\mathrm{i},s}$ the ion-induced secondary electron emission coefficient, and $\Gamma_{\mathrm{i,el},s}$ the ion flux density towards either the right or left electrode surface.
		Whenever a secondary electron is launched from the surface, the respective counter $N_\mathrm{SE,e/l}$ is reduced by one or for fractions reset to zero.
		The coefficients are estimated based on an empirical formula given by Raizer \cite{Raizer1991} that finds application in recent work \cite{Eremin2015, Liu2020}.
		The formula reads
		\begin{align}
			\gamma_{\mathrm{i},s} = 0.016\, \left( \varepsilon_{\mathrm{thr},s} - 2\, \varepsilon_\mathrm{\phi} \right),
		\end{align}
		with $\varepsilon_{\mathrm{thr},s}$ the ionization threshold energy of species $s$ (cf. tab. \ref{tab:chemistry}), and $\varepsilon_\mathrm{\phi}$ the work function of the electrode surface material (cf. Wilson \cite{Wilson1966}).
		For the He/N\textsubscript{2} mixture at hand, the relevant coefficients approximate to $\gamma_\mathrm{i,He\textsuperscript{+}} = 0.25$, $\gamma_\mathrm{i,He\textsubscript{2}\textsuperscript{+}} = 0.25$, and $\gamma_\mathrm{i,N\textsubscript{2}\textsuperscript{+}} = 0.1$.\par
		To account for Penning ionization according to process no. 31 (cf. tab. \ref{tab:chemistry}), a procedure similar to previous work \cite{Donko2018} is adapted.
		It is assumed that for both excitation reactions of helium (cf. reaction no. 2 and 3, tab. \ref{tab:chemistry}), about 50 percent result in the formation of the metastable states He(2\textsuperscript{1}S) and He(2\textsuperscript{3}S), respectively.
		The time constant for diffusion is small compared to the lifetime of the metastables.
		Thus, diffusion is neglected, and a metastable atom is created by saving the position $x_m$ and generation time $t_m$ of the according excitation event.
		Each metastable is assigned with a lifetime
		\begin{align}
			\tau_m = - \frac{1}{k_\mathrm{r}\, n_\mathrm{N\textsubscript{2}}}\, \log(1 - R_\mathrm{01}).
		\end{align}
		$k_\mathrm{r}$ is the reaction rate for the Penning ionization retrieved from \cite{Brok2005, Sakiyama2006}, $n_\mathrm{N\textsubscript{2}}$ denotes the neutral gas density of nitrogen, and $R_\mathrm{01}$ is a pseudo-random number on the interval $[0,1)$.
		For $t \geq t_m + \tau_m$, the metastable decays via the Penning process by contributing to the ionization at position $x_m$.

\newpage


\section{Simulation results and discussion} \label{simulation}

\begin{figure}[t!h!]
		\centering
		\includegraphics[width = \textwidth]{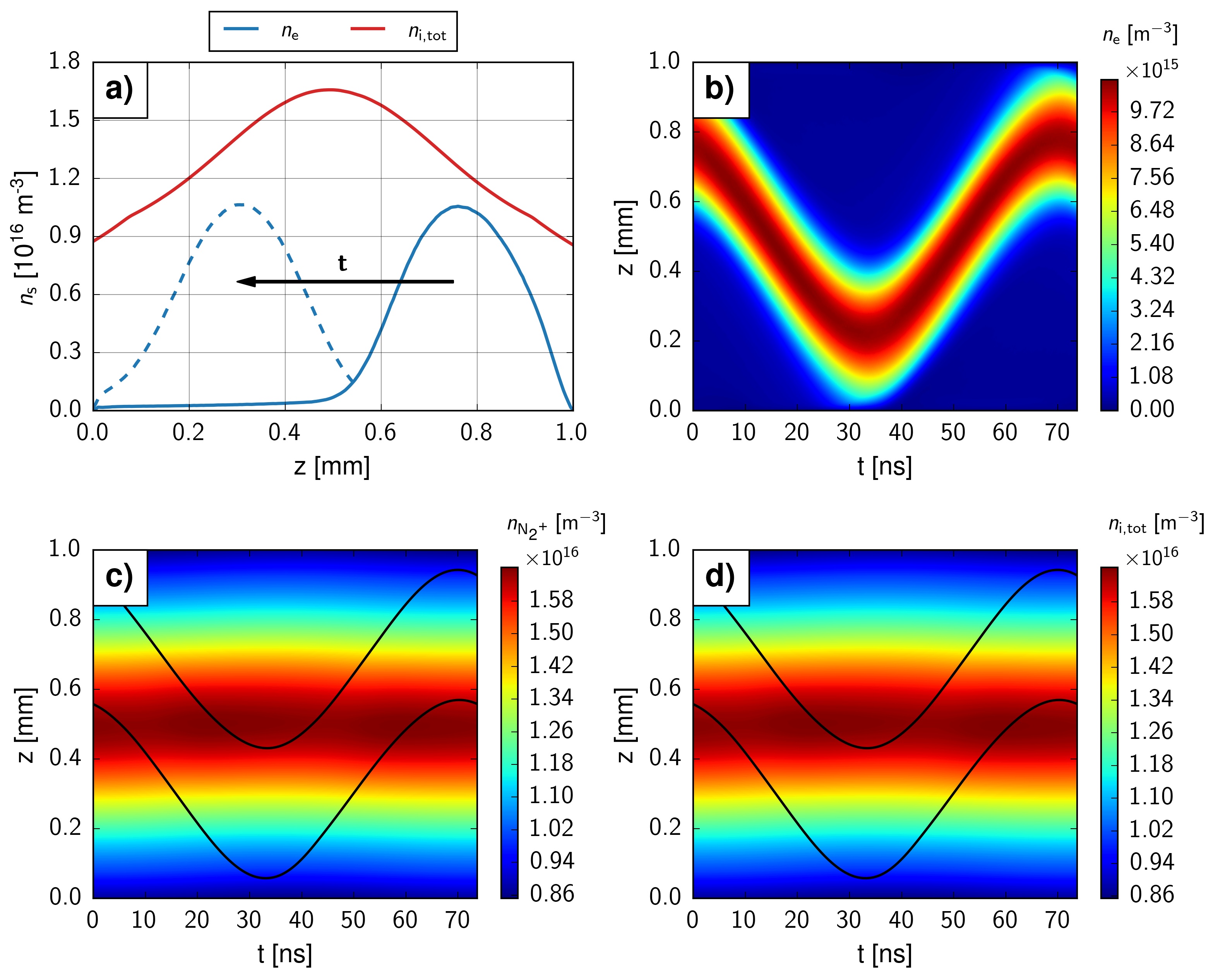}
		\caption{Simulated profiles of the number density $n_s$.
			a): A temporal snapshots of the spatial profile of the electron and total ion density (movie 1 is provided to show the full dynamics).
			b) to d): Temporally and spatially resolved profile of the electron density $n_\mathrm{e}$ (b)), 
		    the N\textsubscript{2}\textsuperscript{+} ion density $n_\mathrm{N\textsubscript{2}\textsuperscript{+}}$ (c)), 
		    and the total ion density $n_\mathrm{i,tot}$ (d)).
			The black lines indicate the estimated position of the electron group.
			The data are obtained from the hybrid PIC/MCC simulation with the parameters: $V_\mathrm{RF} = 220\,$V, $x_\mathrm{N\textsubscript{2}\textsuperscript{+}} = 0.01\,$,  $\gamma_\mathrm{i,He\textsuperscript{+}} = \gamma_\mathrm{i,He\textsubscript{2}\textsuperscript{+}} = 0.25$, $\gamma_\mathrm{i,N\textsubscript{2}\textsuperscript{+}} = 0.1$.}
		\label{fig:XTprofiles}
	\end{figure}

	The most notable characteristics of the non-neutral regime of atmospheric pressure capacitively coupled plasma jets are seen in the density profiles.
	Figure \ref{fig:XTprofiles} a) shows a snapshot of the electron density $n_\mathrm{e}$ (blue), the total ion density $n_\mathrm{i,tot}$ (red) with spatial resolution.
	The panels b) ($n_\mathrm{e}$), c) ($n_\mathrm{N\textsubscript{2}\textsuperscript{+}}$)), and d) ($n_\mathrm{i,tot}$) provide spatially and temporally resolved representations of the respective number densities.
	First, the figure shows that the N\textsubscript{2}\textsuperscript{+} ions are the clearly dominant ion species for this case (fig. \ref{fig:XTprofiles} c), and d)).
	Panels c) and d) show no by naked eye visible difference.
	The density of He\textsubscript{2}\textsuperscript{+} ions already is about three orders of magnitude lower than that of the N\textsubscript{2}\textsuperscript{+} ions.
	For this case, the density of He\textsuperscript{+} ions, which is even lower than the density of the He\textsubscript{2}\textsuperscript{+} ions.
	Hence, both ion densities are not shown.
	The extremely low He\textsuperscript{+} ion density stems from the efficient ion conversion He\textsuperscript{+} $\to$ He\textsubscript{2}\textsuperscript{+} (c.f. reaction 32 tab. \ref{tab:chemistry}).
	This result is in accordance to Martens et al. \cite{Martens2008}.
	Both the He\textsubscript{2}\textsuperscript{+} and the N\textsubscript{2}\textsuperscript{+} ion density profiles exhibit only a slight temporal modulation.
	The plasma is operated in the RF regime ($\omega_\mathrm{p,i} \ll \omega_\mathrm{RF} \ll \omega_\mathrm{p,e}$).\par
	For electrons, on the other hand, figures \ref{fig:XTprofiles} a) and b) show a highly time-dependent structure.
	The spatial profile of the electron density $n_\mathrm{e}$ is approximately Gaussian-shaped (fig. \ref{fig:XTprofiles} a)).
	The temporally and spatially resolved profile shows a narrow band of electrons pushed back and forth between the electrodes while approximately maintaining their mold (fig. \ref{fig:XTprofiles} b)).
	This behavior reminds us on two of the three key characteristics of solitons as given by Drazin and Johnson (\cite{Drazin1989} p. 15): (i) a soliton has a permanent shape, (ii) a soliton is localized within a specific region, and (iii) solitons can interact without changing their shape.
	The dynamics of the electron group of figure \ref{fig:XTprofiles} clearly meets the first two criteria.
	The third criterion cannot be met by a group of electrons.
	If, for some reason, two of these electron groups existed and met, they would interact and eventually form one bigger group of electrons.
	Hence, the group of electrons is not a real drift-soliton.
	However, we decide to refer to this structure as a drift-soliton in a figurative way. \par 
	The elementary characteristic of the non-neutral operation regime is visible by simultaneously looking at the electron density and ion densities.
	The peak electron density locally never resembles the total ion density.
	Quasi-neutrality is violated along the whole discharge region.
	This non-neutral behavior of the discharge has two implications.
	First, the electron dynamics feature specific characteristics that will be discussed in the following.
	Second, the classical segregation of capacitively coupled plasmas in a sheath and a bulk region is rendered meaningless for this operation regime.
	The concept of the boundary sheath edge, especially, and the analysis of its dynamics that in other studies provide orientation and valuable insight \cite{Korolov2019, Korolov2020} cannot be used in the context of a non-neutral regime.\par
	
	The black lines in figure \ref{fig:XTprofiles} c) and d) are introduced to compensate this loss of reference.
	They refer to the rough position of the soliton-like structure in space and time.
	The shape of the electron group is assumed to be Gaussian to calculate the black lines.
	Based on this assumption, the position of the symmetry axis $Z$ is calculated as the first spatial moment of the electron density, and the width of the structure $\delta_z$ calculates from the second spatial moment.
	The black lines mark the interval $Z \pm 1.28\, \delta_z$, in which 90 percent of the electrons are located.
	
	\begin{figure}[t!]
		\centering
		\includegraphics[width = \textwidth]{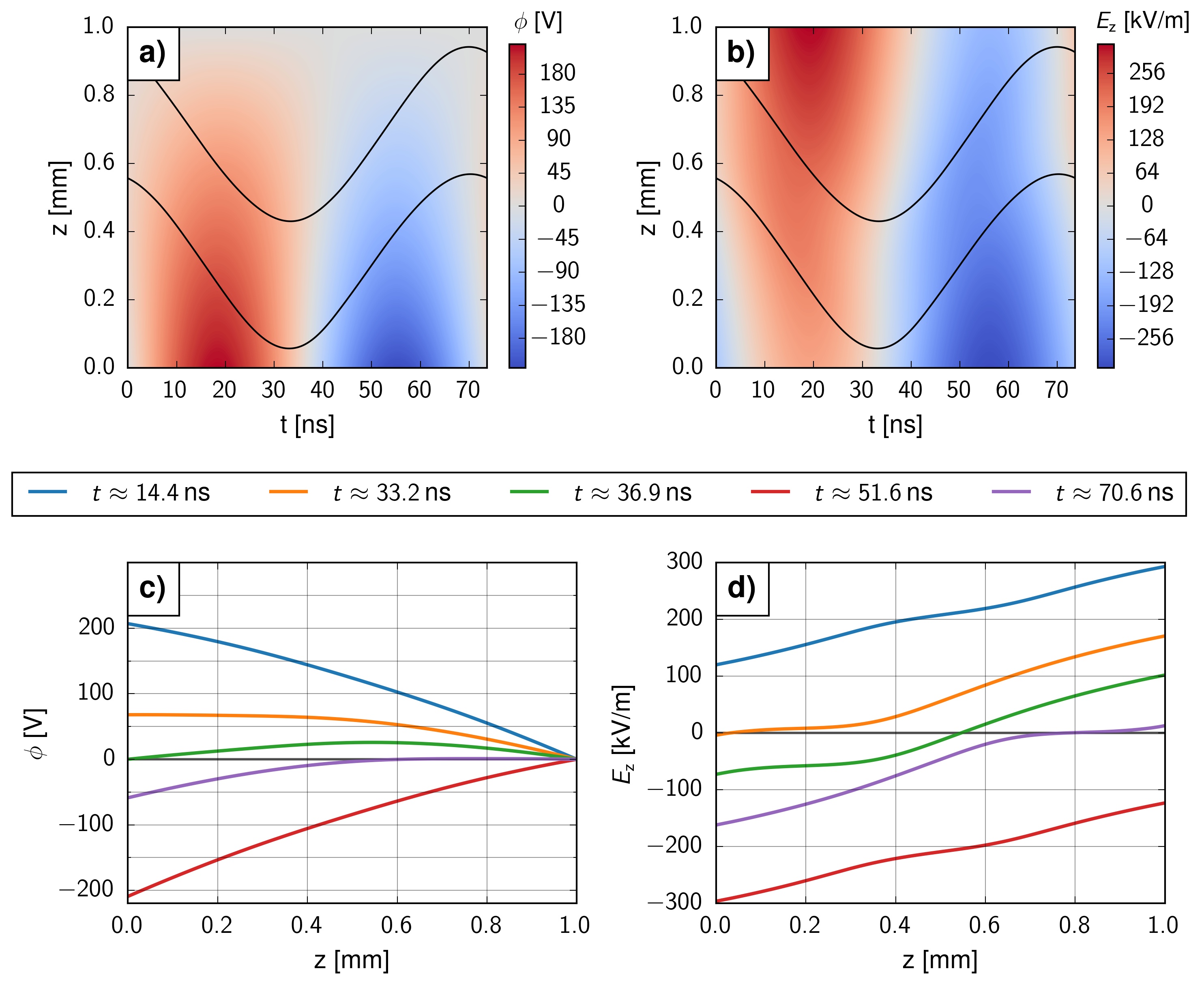}
		\caption{Simulated profiles of the electric potential $\phi$ and the electric field $E_z$.
			a) and b): Temporally and spatially resolved profiles of the potential $\phi$ (a)) and electric field $E_z$ (b)).
			The black lines indicate the estimated position of the electron group.
			c) and d): Characteristic temporal snapshots of the spatial profile of the potential $\phi$ (c)) and the electric field $E_z$ (d)).
			Movies 2 provides an animation of the field dynamics as sketched in panels c) and d).
			The data are obtained from the hybrid PIC/MCC simulation with the parameters: $V_\mathrm{RF} = 220\,$V, $x_\mathrm{N\textsubscript{2}\textsuperscript{+}} = 0.01\,$,  $\gamma_\mathrm{i,He\textsuperscript{+}} = \gamma_\mathrm{i,He\textsubscript{2}\textsuperscript{+}} = 0.25$, $\gamma_\mathrm{i,N\textsubscript{2}\textsuperscript{+}} = 0.1$.}
		\label{fig:XTfields}
	\end{figure}
	
	The non-neutral behavior of the discharge reflects in the electric field and potential.
	Figure \ref{fig:XTfields} shows the electric potential $\phi$ (a)) and the electric field in $z$-direction $E_z$  (b)) with spatial and temporal resolution.
	The remaining panels show temporal snapshots of the individual profiles with spatial resolution ($\phi$: fig. \ref{fig:XTfields} c), $E_z$: fig. \ref{fig:XTfields} d)).
	Due to not having a quasi-neutral region, the potential of the discharge is, in first approximation, the electrical potential of a parallel plate capacitor bearing a positive space charge \cite{JacksonBook}.
	Therefore, the spatial profile of the electric potential resembles a compressed parabola (fig. \ref{fig:XTfields} c)), and the maxima and minima of the potential are primarily located at the electrodes.
	Accordingly, the electric field $E_z$ does not show the bulk-specific plateau around zero.
	When the electron group is around in the middle of the discharge (e. g. fig. \ref{fig:XTfields} d), blue and red lines), the spatial profile of the electric field becomes almost linear (with a small plateau at the position of the maximal electron density).
	For the brief moments when the soliton-like structure is closest to either electrode, a tiny area develops that is closest to being quasi-neutral.
	A corresponding plateau close to the respective electrode becomes visible (e.g., fig. \ref{fig:XTfields} d), orange and purple lines).
	Additionally, and in contrast to a quasi-neutral regime, the electric field often has a single direction and does not cross zero (fig. \ref{fig:XTfields} b) and d)).
	
	\begin{figure}[t!]
		\centering
		\includegraphics[width = \textwidth]{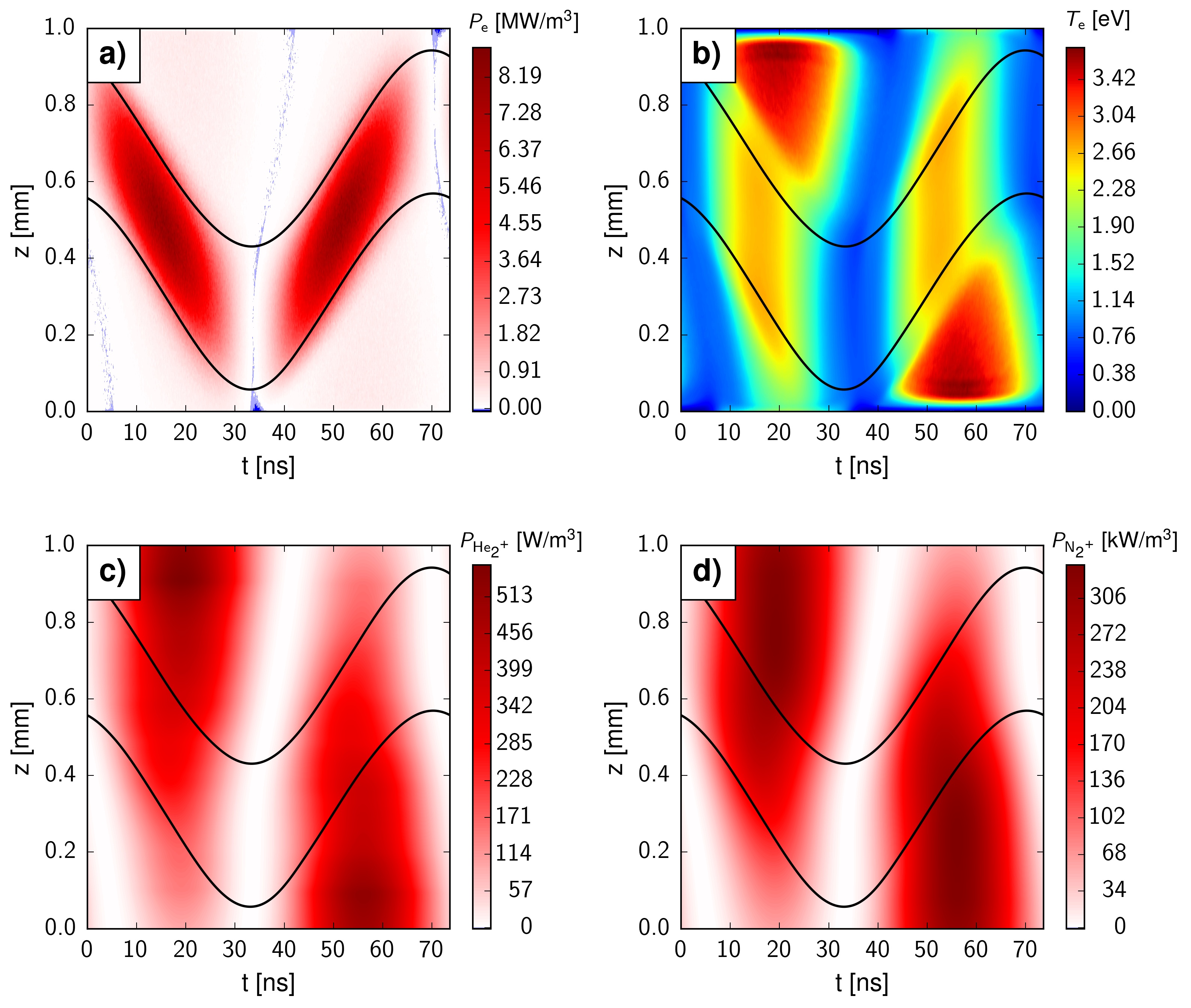}
		\caption{Simulated temporally and spatially resolved profiles of
			a) the power density absorbed by electrons $P_\mathrm{e}$,
			b) the electron temperature $T_\mathrm{e}$,
			c) the power density absorbed by He\textsubscript{2}\textsuperscript{+} ions $P_\mathrm{He\textsubscript{2}\textsuperscript{+}}$, 
			and d) the power density absorbed by N\textsubscript{2}\textsuperscript{+} ions $P_\mathrm{N\textsubscript{2}\textsuperscript{+}}$.
			The black lines indicate the estimated position of the electron group.
			The data are obtained from the hybrid PIC/MCC simulation with the parameters: $V_\mathrm{RF} = 220\,$V, $x_\mathrm{N\textsubscript{2}\textsuperscript{+}} = 0.01\,$,  $\gamma_\mathrm{i,He\textsuperscript{+}} = \gamma_\mathrm{i,He\textsubscript{2}\textsuperscript{+}} = 0.25$, $\gamma_\mathrm{i,N\textsubscript{2}\textsuperscript{+}} = 0.1$.}
		\label{fig:dynamics}
	\end{figure}
	
	The, in the context of a capacitively coupled plasma, unique electric field structure and the strict localization of the electrons within a soliton-like group reflect in the whole electron dynamics.
	Figure \ref{fig:dynamics} shows the power density $P_s$ for different particle species (electrons: a), He\textsubscript{2}\textsuperscript{+} ions: c), N\textsubscript{2}\textsuperscript{+} ions: d)) and the electron temperature $T_\mathrm{e}$ (b)).
	The species-specific power density $P_s$ is calculated as the product of the individual current density $j_s$ and the electrical field $E_z$.
	We retrieve $T_\mathrm{e}$ from the diagonal element $p_\mathrm{zz}$ of the pressure tensor, which is defined as
	\begin{align}
		p_\mathrm{zz} = m_\mathrm{e}\, n_\mathrm{e}\, \left( \langle v_z^2 \rangle - u^2 \right)
	\end{align}
	with $m_\mathrm{e}$ the electron mass, $n_\mathrm{e}$ the electron number density, $\langle v_z^2 \rangle$ the electron's mean-square velocity, and $u$ the mean velocity of the electrons.
	Details of these diagnostics are found in previous work \cite{Wilczek2020}. \par
	The dynamics of the soliton-like structure are reflected in the electron power absorption patterns.
	As discussed before, 90 percent of the electrons are located inside the ``soliton area'' enclosed by the black curves (fig. \ref{fig:dynamics} a)).
	Accordingly, energy loss and gain almost exclusively happen inside this region.
	Whenever the structure moves, electrons are accelerated along its path and gain energy from interacting with the electrical field.
	The most substantial energy gain is observed when the electron group approaches the middle of the discharge gap, and the movement speed of the group directly correlates to the energy gain.
	While changing its direction at the electrodes, the soliton-like structure becomes decelerated and accelerated in the opposing direction afterward.
	The described behavior is, for example, seen between $25$ and $45\,$ns (fig. \ref{fig:dynamics} a)).
	Furthermore, the moments when the movement speed of the electron group drops to zero ($t \approx 35\,$ns and $t \approx 70\,$ns) are the sole opportunities for a significant amount of electrons to leave the discharge.
	At these times, the soliton-like structure is closest to the electrodes, and the loss of electrons coincides with the only visible net loss of energy.\par
	In summary, the electron dynamics of the non-neutral regime are coupled to the soliton-like structure rather than the oscillating boundary sheath (that in this case -at least in its classical sense- does not even exist).
	As a result, energy gain patterns along the movement of the electron group outweigh energy loss patterns at a standstill by a lot.
	Additionally, the energy absorbed by electrons (fig. \ref{fig:dynamics} a)) completely overshadows the energy delivered to the ions (fig. \ref{fig:dynamics} c) and d)).
	The power absorption by ions (fig. \ref{fig:dynamics} c) and d)) basically mimics the particular structure of the electrical field (fig. \ref{fig:XTfields} b)).
	For the positive half-wave of the RF voltage ($t \approx 0 - 37\,$ns), the electrical field is approximatively positive over the whole discharge gap, and ions are pushed towards the grounded electrode.
	For the negative half-wave of the RF voltage ($t \approx 37 - 74\,$ns), the situation is reversed.
	Ions are pulled towards the powered electrode.
	The energy absorption linked to these dynamics is simple.
	Both He\textsubscript{2}\textsuperscript{+} and N\textsubscript{2}\textsuperscript{+} ions show a similar absorption pattern.
	The stronger the electric field, the more energy is absorbed.
	Accordingly, the maxima of the absorbed power are in front of the electrodes and coincide with the maxima of the electrical field (comp. fig. \ref{fig:XTfields} b) to fig. \ref{fig:dynamics} c) and d)).\par
	Another impact of the field structure shows inside the temporal and spatial behavior of the electron temperature displayed in figure \ref{fig:dynamics} b).
	The figure, first of all shows, that the majority of the electrons (the ones inside the ``soliton area'') share the same temperature.
	Outside of the ``soliton area'', structures are visible that correlate to the secondary electron emission.
	The maxima of the electron temperature are located in front of the electrodes and switch sides each half-wave.
	These maxima coincide with the maxima of the electrical field (fig. \ref{fig:XTfields} b)) and the maxima of the ion energy absorption (fig. \ref{fig:dynamics} c) and d)).
	At $t \approx 20\,$ns (grounded) and $t \approx 55\,$ns (powered), the amount of ions reaching the respective electrode is highest, and so is the production of secondary electrons.
	Simultaneously, the absolute value of the field strength is maximal, and the secondary electrons gain an equivalent amount of energy.\par
	Nevertheless, figure \ref{fig:dynamics} b) is misleading regarding the role of secondary electrons.
	First, their number is low.
	Recalling the meaning of the black curves, it becomes evident that a tiny fraction of a few percent of the electron population consists of secondary electrons.
	Second, secondary electrons absorb an insignificant amount of the energy.
	Figure \ref{fig:dynamics} a) shows structures (in light red color) in front of the respective electrodes (grounded: positive half-wave, powered: negative half-wave).
	However, the simultaneous energy gain by electrons inside the soliton-like structure outweighs these patterns.
	Third, another kind of electrons, the ones created by Penning ionization, is more crucial to the non-neutral regime.
	An investigation of the ionization dynamics will prove this point.
		
	\begin{figure}[t!]
		\centering
		\includegraphics[width = \textwidth]{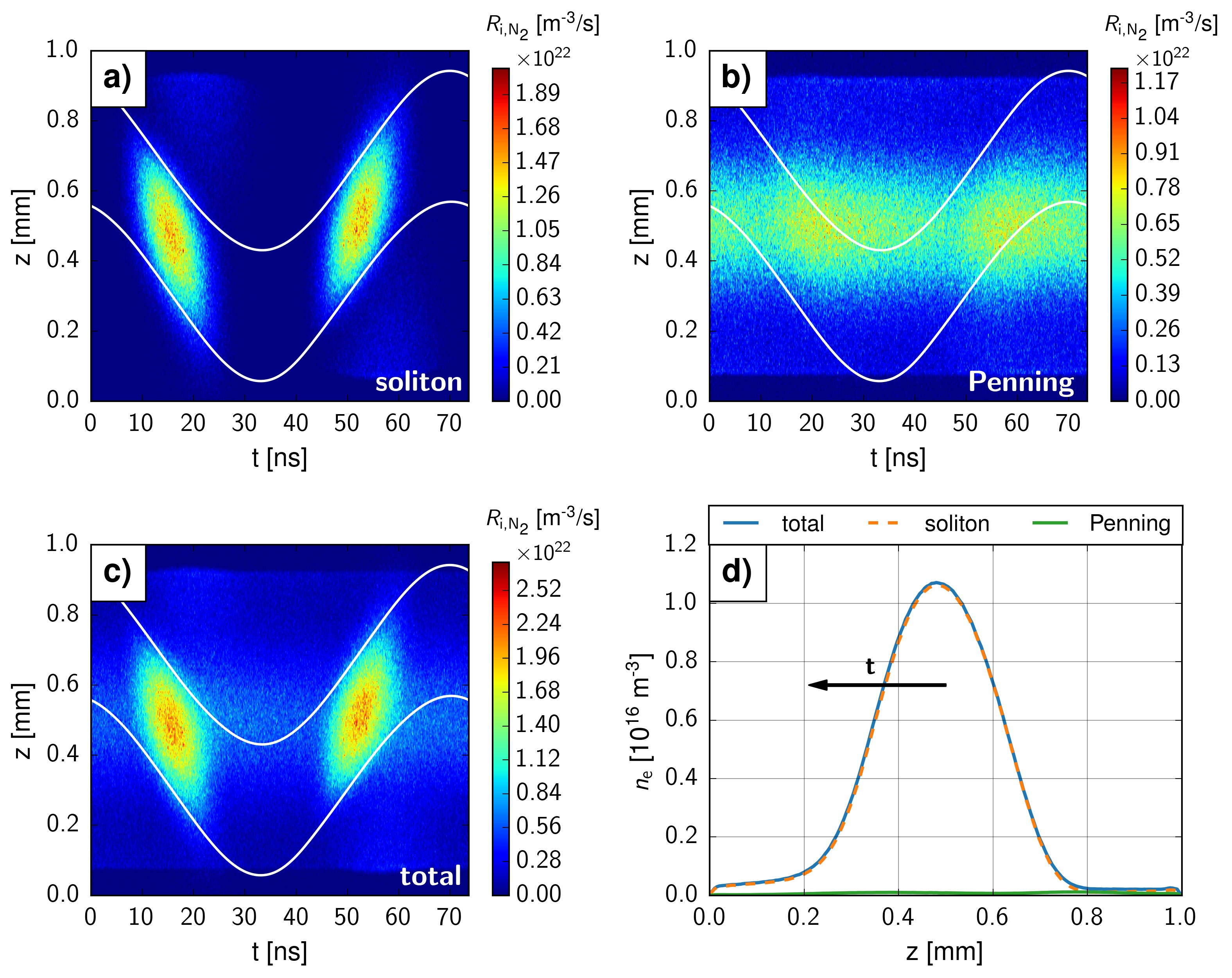}
		\caption{a) to c): Temporally and spatially resolved profiles of the ionization dynamics of N\textsubscript{2}.
			The panels show the ionization rate $R_\mathrm{i,N\textsubscript{2}}$ subdivided into
			a) ionization caused by electrons associated with the soliton-like structure,
			b) ionization due to Penning ionization and the first ionization process of a ``Penning electron'',
			c) the total ionization rate considering all sources of ionization.
			We refer to the text for a detailed explanation of the division.
			d) shows a characteristic temporal snapshot of the spatial profile of the density of electron species classified in the same manner as the ionization (movie 3 provides an animated version of this diagnostics).
			The white lines indicate the estimated position of the electron group.
			The data are obtained from the hybrid PIC/MCC simulation with the parameters: $V_\mathrm{RF} = 220\,$V, $x_\mathrm{N\textsubscript{2}\textsuperscript{+}} = 0.01\,$,  $\gamma_\mathrm{i,He\textsuperscript{+}} = \gamma_\mathrm{i,He\textsubscript{2}\textsuperscript{+}} = 0.25$, $\gamma_\mathrm{i,N\textsubscript{2}\textsuperscript{+}} = 0.1$.}
		\label{fig:ionization}
	\end{figure}
	
	First, the term ``Penning electron'' must be defined.
	For this study, we chose to define an electron as a ``Penning electron'' when created by the Penning ionization process (tab. \ref{tab:chemistry}: No. 31).
	Penning electrons keep their status until their first ionizing collision.
	Please note that this definition is somehow arbitrary.
	According to the above definition, an electron avalanche or ionization cascade started by a ``Penning electron'' will be missed.
	Therefore, the importance of Penning ionization may be underestimated by our means.
	As the following section will point out, this source of arbitrariness cannot affect our results, and any representation of different ionization channels is sufficient for our arguments.\par
	Panels a) to c) in figure \ref{fig:ionization} present temporally and spatially resolved profiles of the ionization rate of nitrogen differentiated by the diagnostics as mentioned above.
	The ionization patterns of the electrons inside the soliton-like structure (fig. \ref{fig:ionization} a)) coincide with the patterns of electron power absorption (fig. \ref{fig:dynamics} a)).
	As discussed before, electrons mainly gain energy while the electron group as a whole is moving.
	Accordingly, there is ionization during the same time.
	When the electron group stops (at powered electrode: $t \approx 35\,$ns, at grounded electrode: $t \approx 70\,$ns), electrons get lost, lose energy, and do not ionize.
	The light green structures adjacent to the respective electrodes (grounded: $t \approx 10 - 25\,$ns, powered: $t \approx 50 - 65\,$ns) stem from ionization associated with the secondary electrons.
	Their contribution is insignificant compared to the ionization inside the soliton-like structure and the ionization pattern of Penning electrons.\par
	The latter group of electrons has a diffuse temporal structure (fig. \ref{fig:ionization} b)).
	Due to its nature and the high pressure, Penning ionization introduces a delay between the initial electron impact producing the metastable helium state and the actual ionization when the metastable decays via collision with a nitrogen molecule.
	As a result, the ionization pattern of Penning electrons is vaguely connected to the energy absorption (the more energy, the more likely an inelastic collision).
	However, the distinct temporal structure becomes smeared out.
	The remainder is a spatial profile with an ionization maximum between $z = 0.4 - 0.6\,$mm.
	Nonetheless, Penning ionization happens always and anywhere among the discharge gap (except a small region in front of the electrodes where no high energetic electrons are present at all).\par
	In the total ionization pattern (fig. \ref{fig:ionization} c)), the structure caused by the movement of the electron group is dominant.
	Nevertheless, the ionization band in the region $z = 0.4 $ to $0.6\,$mm caused by Penning ionization is still visible.
	This observation and the comparable scales of figures \ref{fig:ionization} a) and b) lead to the conclusion that Penning ionization has a significant share on the total ionization.
	Anyhow, figure \ref{fig:ionization} d) shows that the density made up by Penning electrons is insignificant, and there is no sign of different behavior of the Penning electrons.
	Apparently, electrons are rapidly incorporated into the soliton-like struc\-ture regardless of their origin.
	The discussion of figure \ref{fig:dynamics} b) regarding the electron temperature $T_\mathrm{e}$ painted a similar picture.
	All electrons more or less share the same fate, and neither diagnostic nor analysis of the simulation data could provide an answer.
	For now, two key questions remain open: (i) How does the soliton-like structure form inside the non-neutral regime?, and (ii) Why do the electrons behave similarly? \par 

\pagebreak


\section{Mathematical modeling} \label{modeling}

To develop a better physical understanding of the discharge dynamics and to answer the   questions raised above, we now analyze an elementary 
drift-diffusion model.
\linebreak
Such models trade, in a way, quantitative accuracy for mathematical transparency \cite{VanDijk2009}.
We are interested in the behavior of a soliton (the term used in the sense of section \ref{simulation}) that is not disturbed by the boundaries, and
extend the solution domain to the real axis.
\linebreak
Moreover, we focus on the dynamics of the electrons and the electric field and treat the ion density $n_\mathrm{i}$ 
as a spatial and temporal constant.
The electron model is given by the equation of continuity with the generation term neglected,
\begin{align}
    \frac{\partial n_\mathrm{e}}{\partial t} + \frac{\partial \Gamma_\mathrm{e}}{\partial z} =0. \label{eq:DD3} 
\end{align}
The flux $\Gamma_\mathrm{e}$ is calculated with the diffusion constant $D_\mathrm{e}$ and the mobility $\mu_\mathrm{e}$,
\begin{align}
	  \Gamma_\mathrm{e} &= -D_\mathrm{e}\, \frac{\partial n_\mathrm{e}}{\partial z} -\mu_\mathrm{e}\, E_z\, n_\mathrm{e}. \label{eq:DD4}
\end{align}
The electric field $E_z(z,t)$ and its potential $\phi(z,t)$ are given by Poisson's equation:
\begin{align}
	\varepsilon_0\, \frac{\partial E_z}{\partial z}= - \varepsilon_0\, \frac{\partial^2 \phi}{\partial z^2} = e\, (n_\mathrm{i} - n_\mathrm{e}). \label{eq:Poisson}
\end{align}
The total discharge current density $j_z(t)$ is assumed to be periodic in time, average-free and spatially homogeneous. It is related to the surface charge density,  $Q(t)$, as \cite{Brinkmann2009}:
\begin{align}
 - e\, \Gamma_\mathrm{e} + 		\varepsilon_0\, \frac{\partial E_z}{\partial t}
 = j_z(t) = -\frac{dQ}{dt}. \label{eq:current}
\end{align}
The system of equations is completed by suitable asymptotic conditions for $z \to \pm \infty$. The electron
density is assumed to vanish and the electric field is asymptotically linear;
the constants $E_{\pm \infty}$ have the dimension of the electric field:
	\begin{align}
	   & n_\mathrm{e} \big|_{z \to \pm \infty} = 0,\\[0.25ex]
	   & E_z  \,  - \frac{1}{\varepsilon_0}\,e\, n_\mathrm{i}\,z\Big|_{z \to \pm \infty} = 
	   - \frac{1}{\varepsilon_0}\, Q(t) + E_{\pm \infty}.
		\end{align}
	
The global dynamics of the electron soliton can be described by a closed model. 
We define three spatial moments of the electron density, namely
the total number $\mathcal{N}(t)$, the center of mass (COM) $\mathcal{Z}(t)$,
and the soliton-averaged field $\mathcal{E}(t)$:
\begin{align}
    \mathcal{N}(t) &= \int_{- \infty}^\infty n_\mathrm{e}(z,t)\, \mathrm{d} z \label{eq:NE},\\
    \mathcal{Z}(t) &= \frac{1}{\mathcal{N}(t)}\int_{- \infty}^\infty z\, n_\mathrm{e}(z,t)\, \mathrm{d} z \label{eq:Z},\\
    \mathcal{E}(t) &= \frac{1}{\mathcal{N}(t)}\int_{- \infty}^\infty E_z(z,t)\, n_\mathrm{e}(z,t)\, \mathrm{d} z \label{eq:Eeff}.
\end{align}

Integration of the particle balance equation \eqref{eq:DD3} over the spatial axis 
leads to the insight that the particle number $\mathcal{N}$ is constant.
Applying the same integral to Poisson's equation \eqref{eq:Poisson} results in
an identity which states that the offsets in the asymptotically linear 
electric field forms are related to the electron number:
\begin{align}
	\mathcal{N} = -\frac{\varepsilon_0}{e}\, \left( E_{+\infty} - E_{-\infty} \right).
\end{align}
Integrating the continuity equation with the weight $z$, we derive a differential equation that describes the drift motion of the COM with regard to the average field:
		\begin{align}
			\frac{\mathrm{d}{\mathcal{Z}}}{\mathrm{d}{t}} = - \mu_\mathrm{e}\, \mathcal{E}. \label{eq:Z1}
		\end{align}
Further, by manipulating the model equations and utilizing the asymptotic conditions, 
we~obtain two additional identities. The first identity states
\begin{align}
	\mathcal{E} = \frac{e\, n_\mathrm{i}}{\varepsilon_0}\, \mathcal{Z} - \frac{1}{\varepsilon_0}\, Q(t) + \frac{E_{+\infty} + E_{- \infty}}{2}. \label{eq:Z2}
\end{align}
The second identity directly relates the current and the density averaged field:
\begin{align}
  \varepsilon_0\frac{\mathrm{d}\mathcal{E}}{\mathrm{d}t}+\mu_\mathrm{e}  e n_\mathrm{i} \mathcal{E}  = j_z
\end{align}
Equations \eqref{eq:Z1} and \eqref{eq:Z2} combine to a closed 
differential equation for the COM:
\begin{align}
	\frac{\mathrm{d}\mathcal{Z}}{\mathrm{d}t} = \frac{\mu_\mathrm{e}}{\varepsilon_0}\, Q(t) - \frac{e\, n_\mathrm{i}\, \mu_\mathrm{e}}{\varepsilon_0}\, \mathcal{Z} - \mu_\mathrm{e}\, \frac{E_{+ \infty} + E_{- \infty}}{2}. \label{eq:dglZ}
		\end{align}
When viewed as an initial value problem, the solution $\mathcal{Z}(t)$ of this differential equation depends on the initial value $\mathcal{Z}_0$ at $t_0$. 
Asymptotically, however, a periodic solution $ \mathcal{Z}_\mathrm{as}(t)$
is reached which solely depends on the modulation $Q(t)$ and the asymptotic conditions. The first term is constant and the
second term is periodic and average-free:
\begin{align}
    \mathcal{Z}_\mathrm{as}(t) = - \frac{\varepsilon_0}{2\, e\, n_\mathrm{i}}\, \left( E_{+ \infty} + E_{- \infty} \right) + \frac{\mu_\mathrm{e}  }{\varepsilon_0} \int_{-\infty}^{\mathrm{t}} \exp\left(-\frac{\mathrm{e} n_\mathrm{i}  \mu_\mathrm{e}}{\varepsilon_0}(t-\tau)\right) Q(\tau)\,  \mathrm{d} \tau.  
\end{align}
The corresponding density averaged field is
\begin{align}
	\mathcal{E}_\mathrm{as}(t) = \frac{e\, n_\mathrm{i}}{\varepsilon_0}\, \mathcal{Z}_\mathrm{as} - \frac{1}{\varepsilon_0}\, Q(t) + \frac{E_\infty + E_{- \infty}}{2}. \label{eq:Z2As}
\end{align}
This point $ \mathcal{Z}_\mathrm{as}(t)$ is now taken as the origin of 
new coordinates which we, for a reason that will become clear shortly,
call the current-free system:
\begin{align}
     z = \mathcal{Z}_\mathrm{as}(t) + \xi. 
\end{align}
The electron density, the electron flux, and the electric field are transformed as follows, where $n_\mathrm{e}^\prime(\xi, t)$, $\Gamma_\mathrm{e}^\prime(\xi, t)$, and $E_z^\prime(\xi, t)$ are measured in the new coordinates:
\begin{align}
     n_\mathrm{e} (z,t) &= n_\mathrm{e}^\prime (\xi, t), \\
     \Gamma_\mathrm{e} (z,t) &= n_\mathrm{e}^\prime (\xi, t)\, \frac{\mathrm{d} \mathcal{Z}_\mathrm{as}}{\mathrm{d}t}+\Gamma_\mathrm{e}^\prime (\xi, t),\\
      E_z(z,t)&= \mathcal{E}_\mathrm{as}(t) + 	E_z^\prime (\xi,t). 
\end{align}
Immediately dropping the prime, the original equations \eqref{eq:DD3} to \eqref{eq:Poisson} remain unchanged.
For further streamlining, \linebreak we adopt 
a dimensionless notation. We introduce the Einstein temperature $T_\mathrm{e} = e D_\mathrm{e} / \mu_\mathrm{e}$, the Debye length $\lambda_\mathrm{D} = \sqrt{\varepsilon_0 T_\mathrm{e} / e^2 n_\mathrm{i}}$, and the electric relaxation time $\tau_\mathrm{e} = \varepsilon_0 / e \mu_\mathrm{e} n_\mathrm{i}$.
Measuring length in $\lambda_\mathrm{D}$, time in $\tau_\mathrm{e}$,
density in $n_\mathrm{i}$, flux in $n_\mathrm{i} \lambda_\mathrm{D}/\tau_\mathrm{e}$, electric field in $T_\mathrm{e}/e\lambda_\mathrm{D}$,
and electric potential in $T_\mathrm{e}/\lambda_\mathrm{D}$, we obtain:
\begin{align}
	&\frac{\partial n_\mathrm{e}}{\partial{t}}+\frac{\partial \Gamma_\mathrm{e}}{\partial{\xi}} = 0,\label{eq:dimless1} \\
	&\Gamma_\mathrm{e} = - E_z\, n_\mathrm{e} - \frac{\partial{n_\mathrm{e}}}{\partial{\xi}},\label{eq:dimless2}\\
		&\frac{\partial{E_z}}{\partial{\xi}} = - \frac{\partial^2{\phi}}{\partial{\xi}^2} = 1 - n_\mathrm{e},\label{eq:dimless3}\\
	&0 = - \Gamma_\mathrm{e}  +  \frac{\partial{E_z}}{\partial{t}}.\label{eq:dimless4}
\end{align}
The asymptotic conditions translate to
\begin{align}
   & n_\mathrm{e} \big|_{\xi \to \pm \infty} = 0,\\
   & E_z \,  - \xi\big|_{\xi \to \pm \infty}= \pm (E_\infty-E_{-\infty})/2.
\end{align}

\begin{figure}[t!]
			\centering
			\includegraphics[width = \textwidth]{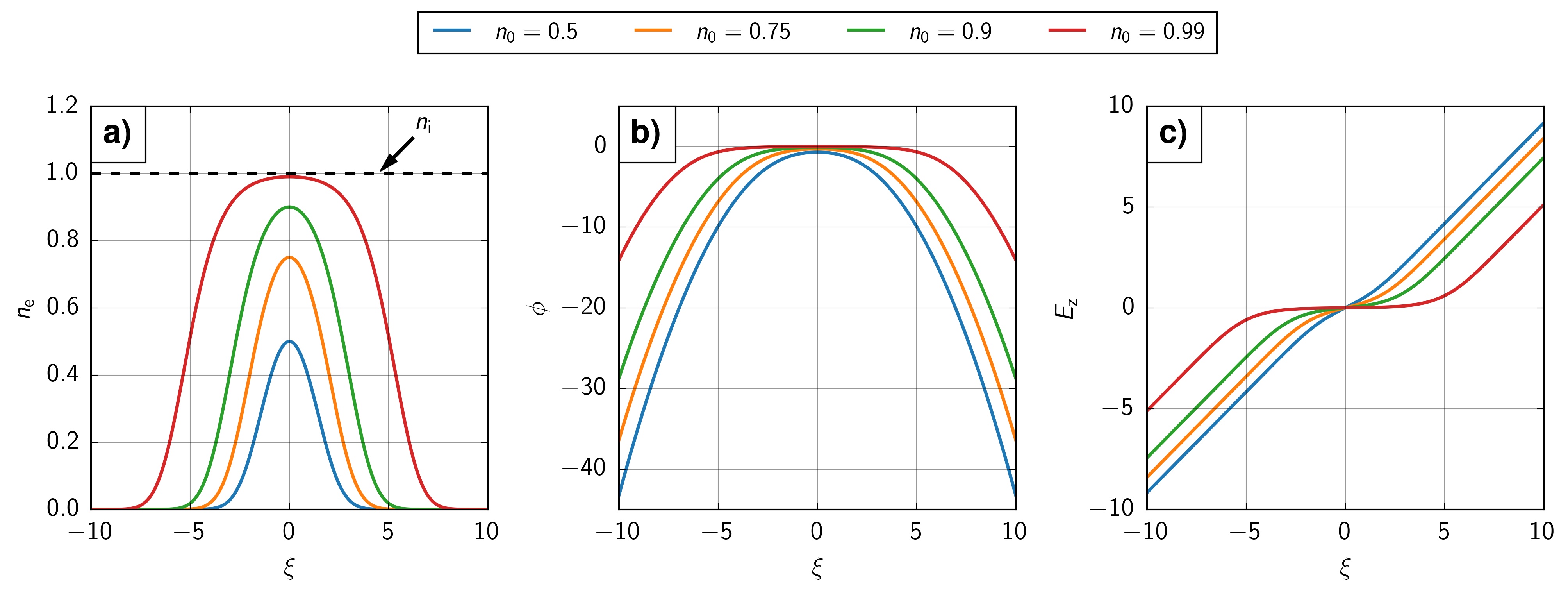}
			\caption{A numerical solution of equation \eqref{eq:DGL} for various parameters $n_0$.
				a) electron density $n_\mathrm{e}$,
				b) electrical potential $\phi$,
				and c) electric field $E_z$.
				All properties are given in terms of dimensionless variables within a current-free reference frame.}
			\label{fig:mod_solutions}
		\end{figure}

In the new coordinates, the equation of motion for the COM is a simple relaxation. The~COM converges exponentially to the reference point 0. The same statement holds for the 
particle-averaged electric field:
\begin{align}
	\frac{\mathrm{d}{\mathcal{Z}}}{\mathrm{d}{t}} &=-\mathcal{Z},\label{eq:dglZCOM} \\
	\frac{\mathrm{d}{\mathcal{E}}}{\mathrm{d}{t}} &=-\mathcal{E}.\label{eq:dglECOM}
\end{align}
Thus, it is  reasonable to assume that the structure as a whole relaxes to an
equilibrium.
The stationary solution of eqs.~\eqref{eq:dimless1} - \eqref{eq:dimless4} has zero flux, obeys Boltzmann equilibrium, 
 and is described by the a second-order differential equation for the potential~$\phi$ known as the Poisson-Boltzmann equation 
 \cite{Maggs2012, Jadhao2013, Lee2014, Gray2018,Brinkmann2007}:
		\begin{align}
			- \frac{\partial^2{\phi}}{\partial{\xi^2}} = 1 - \exp{ \left( \phi \right) }. \label{eq:DGL}
		\end{align}
The Poisson-Boltzmann equation is of second order and therefore needs two conditions.
The symmetry around the origin demands $\partial{\phi} / \partial{\xi} = 0$. The second degree of freedom
can conveniently by fixed by setting $\phi(0) = \ln(n_0)$, with $n_0 \in \left[0,1\right[$.
		This parameter can be identified as the ratio between the electron 
		density and the ion density at $\xi=0$. 
Analytical solutions to \eqref{eq:DGL} are well known but quite inconvenient to handle \cite{Lee2014, Gray2018}. Figure \ref{fig:mod_solutions} presents numerical solutions for the electron density $n_\mathrm{e}$, the electric 
potential~$\phi$, and the electric field $E_z$. The character
of the solutions depends on the parameter $n_0$. \linebreak For $n_0$ close to $1$,
the bulk-sheath structure of low-pressure discharges is recovered.
For values of $n_0$ that are considerably smaller, strongly non-neutral structures appear. 
		Compared to figures \ref{fig:XTprofiles} b) and \ref{fig:XTfields}, the orange and green curves already approximate the previously discussed simulation results (cf. sec. \ref{simulation}).\par
\clearpage
\pagebreak

Intuitively, the stationary solutions are stable. An effective method to show this formally is Lyapunov’s direct method \cite{Lyapunov1966, Lyapunov1992, Pukdeboon2011}. 
The starting point is a balance equation for the density of the Helmholtz free energy \cite{Decyk1982,Ymeri1997, Brinkmann1987, Jadhao2013, Gray2018}:
		\begin{align}
			\frac{\partial}{\partial{t}} \left( \frac{1}{2}\, {E_z}^2 + n_\mathrm{e}\, \log (n_\mathrm{e}) - n_\mathrm{e} \right) +%
			\frac{\partial}{\partial{\xi}} \left( \Gamma_\mathrm{e}\, \log (n_\mathrm{e}) \right) =%
			- \frac{{\Gamma_\mathrm{e}}^2}{n_\mathrm{e}}. \label{eq:Heb}
		\end{align}
		The Helmholtz free energy itself reads as follows, where the temporally constant last term in the integrand serves to render the integral finite:
		\begin{align}
		 \mathcal{F} =   \int_{- \infty}^{\infty}  \Biggl( \frac{1}{2}\,   {E_z}^2    + n_\mathrm{e}\, \log (n_\mathrm{e}) - n_\mathrm{e} - \frac{1}{2}\left( \left| \xi \right| + \frac{\mathcal{N}}{2} \right)^2 \Biggr)  \, \mathrm{d}\xi
		\end{align}
		The time derivative of $\mathcal{F}$ is negative semi-definite, so that it is a suitable candidate for a Lyapunov functional \cite{Brinkmann1987}.
		Note that the equality sign is only assumed when the flux and with it 
		the dissipation vanishes:
		\begin{align}
			\frac{\mathrm{d}\mathcal{F}}{\mathrm{d}t} =-  \int_{- \infty}^{\infty}  	 \frac{{\Gamma_\mathrm{e}}^2}{n_\mathrm{e}}\, \mathrm{d}\xi %
			\leq 0. \label{eq:LyapunovCand}
		\end{align}
		A solution of equation \eqref{eq:DGL} is stable when it represents a minimum of the free energy under the constraint of a fixed particle number $\mathcal{N}$. A constraint-free variational problem can be formulated with the help of a Lagrangian multiplier $\Lambda$:
		\begin{align}
			\mathcal{L} =%
			\mathcal{F} + \Lambda\,\mathcal{N} = %
			\int_{-\infty}^{+ \infty} \left(%
			\frac{1}{2}\, \left( {E_z}^2 - \left( \left| \xi \right| + \frac{\mathcal{N}}{2} \right)^2 \right) + n_\mathrm{e} \bigl( \log (n_\mathrm{e}) - 1 + \Lambda \bigr) %
			\right)\, \mathrm{d}\xi. \label{eq:LagrangianF}
		\end{align}
		The first variation of $\mathcal{L}$ reads as follows, where the second identity is obtained by partial integration of the first term, taking Poisson's equation for $\delta E_z$ into account:
		\begin{align}
		    \delta{\mathcal{L}} =& \int_{- \infty}^{+ \infty} \left( E_z\, \delta{E_z} + \delta{n_\mathrm{e}}\, \log (n_\mathrm{e}) + \Lambda\, \delta{n_\mathrm{e}} \right)\, \mathrm{d}\xi, \label{eq:var1}\\ \nonumber
			   =& %
			\int_{- \infty}^{+ \infty}  
			\delta{n_\mathrm{e}} \bigl( - \phi + \log (n_\mathrm{e}) + \Lambda \bigr)%
		  \, \mathrm{d}\xi. \label{eq:criterium}
		\end{align}
The functional $\mathcal{L}$ has an extremum when $\delta{\mathcal{L}}$ vanishes for all $\delta n_\mathrm{e}$. This implies that the bracket in the second identity vanishes. 
The electrons must be in Boltzmann equilibrium.
The fact that $\Lambda=0$ is a consequence of the choice of units:
\begin{align}
    n_\mathrm{e} = \exp(\phi-\Lambda) \equiv \exp(\phi)
\end{align}
The second variation \eqref{eq:var2} directly shows that all extrema of the functional are minima, the equilibria are therefore stable:
		\begin{align}
			\delta^2{\mathcal{L}} &= \int_{- \infty}^{+ \infty} \Biggl( {\delta{E_z}}^2 + \frac{{\delta{n_\mathrm{e}}}^2}{n_\mathrm{e}} \Biggr) \, \mathrm{d} \xi \geq 0. \label{eq:var2}
		\end{align}

       To assess how quickly the equilibrium is restored, we analyze the evolution of a~small perturbation of a stationary solution of equation \eqref{eq:DGL}.
	   We assume that the perturbation is current-free and has an
	   integrated density of zero. Expressed in the 
	   electrical field $\delta E_z$, \linebreak
	   the evolution can be described by a parabolic 
	   differential equation, completed by suitable 
	   asymptotic conditions and an initial condition at $t=0$:
	 \begin{align}
	       & \frac{\partial \delta E_z}{\partial t}=   \frac{\partial^2{\delta{E_z}}}{\partial{\xi}^2}+  \bar{E}_z \frac{\partial \delta E_z}{\partial \xi}  - \exp\left(\phi\right)  \delta{E_z},\\ \nonumber
		& \delta E_z \big|_{\xi \to \pm \infty} = 0,   \\
		& \delta E_z \big|_{t = 0} = \delta E_0(\xi).   \nonumber 
	   \end{align}

	   It is advantageous to interpret the dynamics in the language of functional analysis. We consider the space of all complex functions $\psi$ on $\mathbb{R}$ which have a finite norm that is motivated by the second variation of the Helmholtz free energy, 
	  \begin{align}
	       \mathbf{H} = 
	       \left\{ \psi \;\Big| \; ||\psi ||^2 =  \int_{-\infty}^\infty \Biggl( 
			|\psi|^2 +
			   \exp\left(-\phi\right)
			   \left|\frac{\partial \psi}{\partial \xi} \right|^2 \Biggr) 
			    \, \mathrm{d}\xi < \infty \right\}.
	  \end{align}
	     This space can be turned into a Hilbert space by defining the scalar product
	    \begin{align}
	        \langle \psi ,\psi^\prime\rangle =  \int_{-\infty}^\infty \Biggl( 
			\psi^* \psi^\prime  +
			   \exp\left(-\phi\right)
			   \frac{\partial \psi^*}{\partial \xi} \,
			   \frac{\partial \psi^\prime}{\partial \xi} 
			   \Biggr)  \, \mathrm{d}\xi.
	    \end{align}
	 The evolution operator is now defined as
	 \begin{align}
	   T \psi  = -\frac{\partial^2{\psi}}{\partial{\xi}^2}-  \bar{E}_z \frac{\partial \psi}{\partial \xi}  + \exp\left(\phi\right)  \psi.
	 \end{align}
	 A short calculation verifies the identity
	 \begin{align}
	     \langle T \psi ,\psi^\prime\rangle
	      =  \int_{-\infty}^\infty \Biggl( 
	      \exp(-\phi) \frac{\partial^2 \psi^*}{\partial \xi^2}\frac{\partial^2 \psi^\prime}{\partial \xi^2}  
	      + \left(3-\exp\left(-\phi\right)\right)\frac{\partial \psi^*}{\partial \xi}\frac{\partial \psi^\prime}{\partial \xi} + \psi^* \psi^\prime 
	      \Biggr) \, \mathrm{d}\xi,
	 \end{align}
	 from which we conclude that the evolution operator is self-adjoint and positive definite, so that its spectrum lies on the positive real axis:
	 \begin{align}
	        \langle T \psi ,\psi^\prime\rangle = 
	             \langle \psi , T \psi^\prime\rangle.
	 \end{align}

\clearpage\newpage	 
	 
	\begin{figure}[t!]
			\centering
			\includegraphics[width = \textwidth]{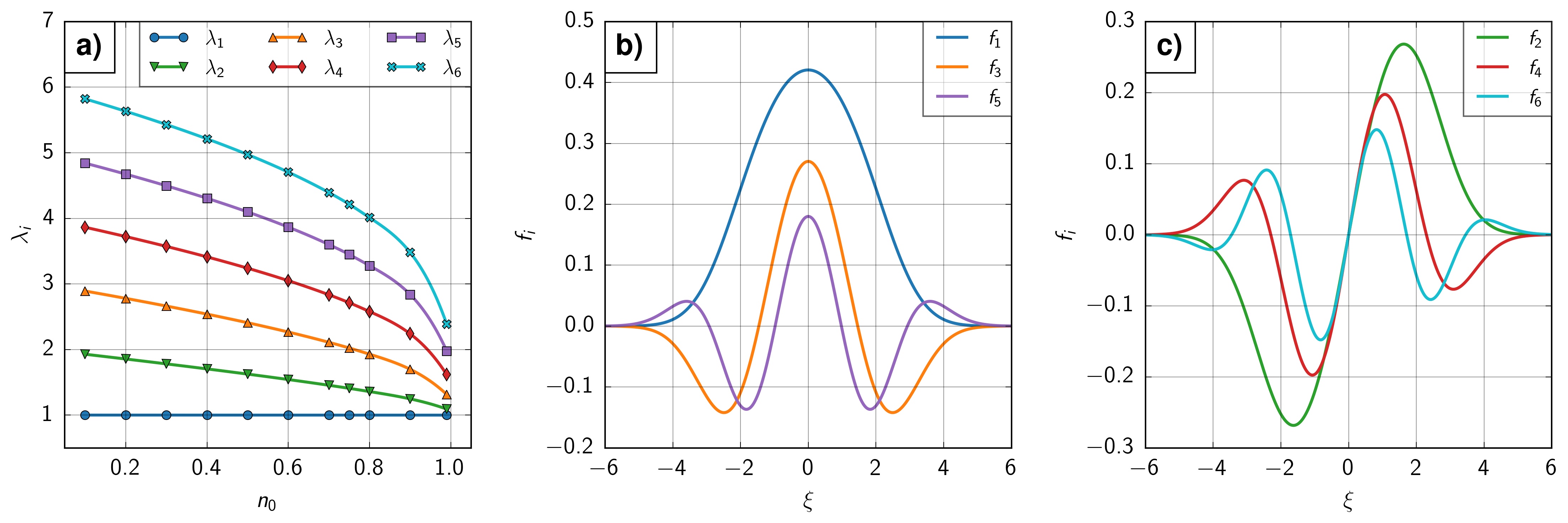}
			\caption{a) The first six numerically calculated eigenvalues of the linear perturbation equation. b)~the~first
			three even eigenfunctions, and c) the first three odd eigenfunctions for $n_0 = 0.75$.}
			\label{fig:eigenvalues}
		\end{figure}
	 
Fig.~\ref{fig:eigenvalues} shows the first eigenvalues and eigenfunctions for the case $n_0 = 0.75$.  
Evidently, one eigenvalue is $\lambda_1 = 1$, the corresponding eigenfunction is
	 \begin{align}
	     \psi_1(\xi)  = \exp\left(\phi(\xi)\right) \sim
	       \frac{\partial \bar{E}_z}{\partial \xi} -1.
	 \end{align}
The following identity (established by means of partial integration) implies that all other eigenvalues are larger than unity. 
	 \begin{align}
	  \langle T \psi ,\psi\rangle  -
	     \langle \psi ,\psi\rangle = & \int_{-\infty}^\infty  \exp(\phi) \left(\frac{\partial^2}{\partial\xi^2}\left(\exp\left(-\phi\right)\psi\right)\right)^2 \, \mathrm{d}\xi \\ \nonumber & + \int_{-\infty}^\infty\exp\left(\phi\right) \left(2-\exp\left(\phi\right)\right)
	   \left(\frac{\partial}{\partial\xi}\left(\exp\left(-\phi\right)\psi\right)\right)^2 \, \mathrm{d}\xi \ge  0.
	 \end{align}

 Written in this compact functional analytic notation, the evolution equation for the perturbation and its initial condition read:
  	 \begin{align}
       & \frac{\partial \delta E_z}{\partial t}+ T\, \delta E_z = 0,\label{eq:dynDGL} \\[0.25ex]
       & \delta E_z \big|_{t = 0} = \delta E_0.  \nonumber
  \end{align}
An application of the Laplace transform yields	 
\begin{align}
    p \underline{\delta E_z}  + T \underline{\delta E_z} =   \delta E_0,
\end{align}
which can be solved in terms of the resolvent: 
\begin{align}
     \underline{\delta E_z}  
    = \left(p +  T\right)^{-1}   \delta E_0,
\end{align}
The Laplace back transform gives an explicit solution:
\begin{align}
    \delta E_z (t) = \int_{-i\infty}^{i\infty} 
       \left(p +  T\right)^{-1} \exp(pt)   \, \mathrm{d}p \, \delta E_0.
\end{align}
Utilizing that all eigenvalues of $T$ are positive, the answer to the 
perturbation is a sum of decaying exponential modes: 
\begin{align}
    \delta E_z(t) = \sum_{n=1}^\infty \langle \Psi_n,\delta E_0 \rangle  \, \exp\left(-\lambda_n t\right)\, \Psi_n.
\end{align}
		
It can be stated, that the system returns to its equilibrium in an exponential fashion.\linebreak The speed is governed by the dielectric relaxation time $\tau_\mathrm{e}$.
For the simulation of section \ref{simulation}, \linebreak
the dielectric relaxation time is $\tau_\mathrm{e} \approx 38.7\,$ns, faster than
the RF period ($T_\mathrm{RF} \approx 73.7\,$ns).
The slowest time constant $\lambda_1= 1$ corresponds to a motion of the
soliton as a whole: If~displaced, the soliton 
returns to its original position within a few relaxation times.
This is already captured by equation \eqref{eq:dglZCOM}. 
The higher modes govern the restoration of the shape and act 
even faster. When distorted, for example by contact with 
the~walls,
the soliton assumes its orginal shape quite quickly.

\clearpage		
\newpage


\section{Simulation and model: Physical conclusions} \label{comparison}
	
The previous sections have described the non-neutral discharge regime in terms of a hybrid PIC/MCC simulation 
(section \ref{simulation}) and a simplified analytical model (section~\ref{modeling}). How do the descriptions compare?
For reference, we extract from our hybrid PIC/MCC simulation
at each time $t$ the ratio $n_0 = n_\mathrm{i}/n_\mathrm{e}$ and calculate the solution of equation~\eqref{eq:DGL}.
All quantities are transformed back into physical units and the laboratory frame.

		\begin{figure}[h!]
		\centering
		\includegraphics[width = \textwidth]{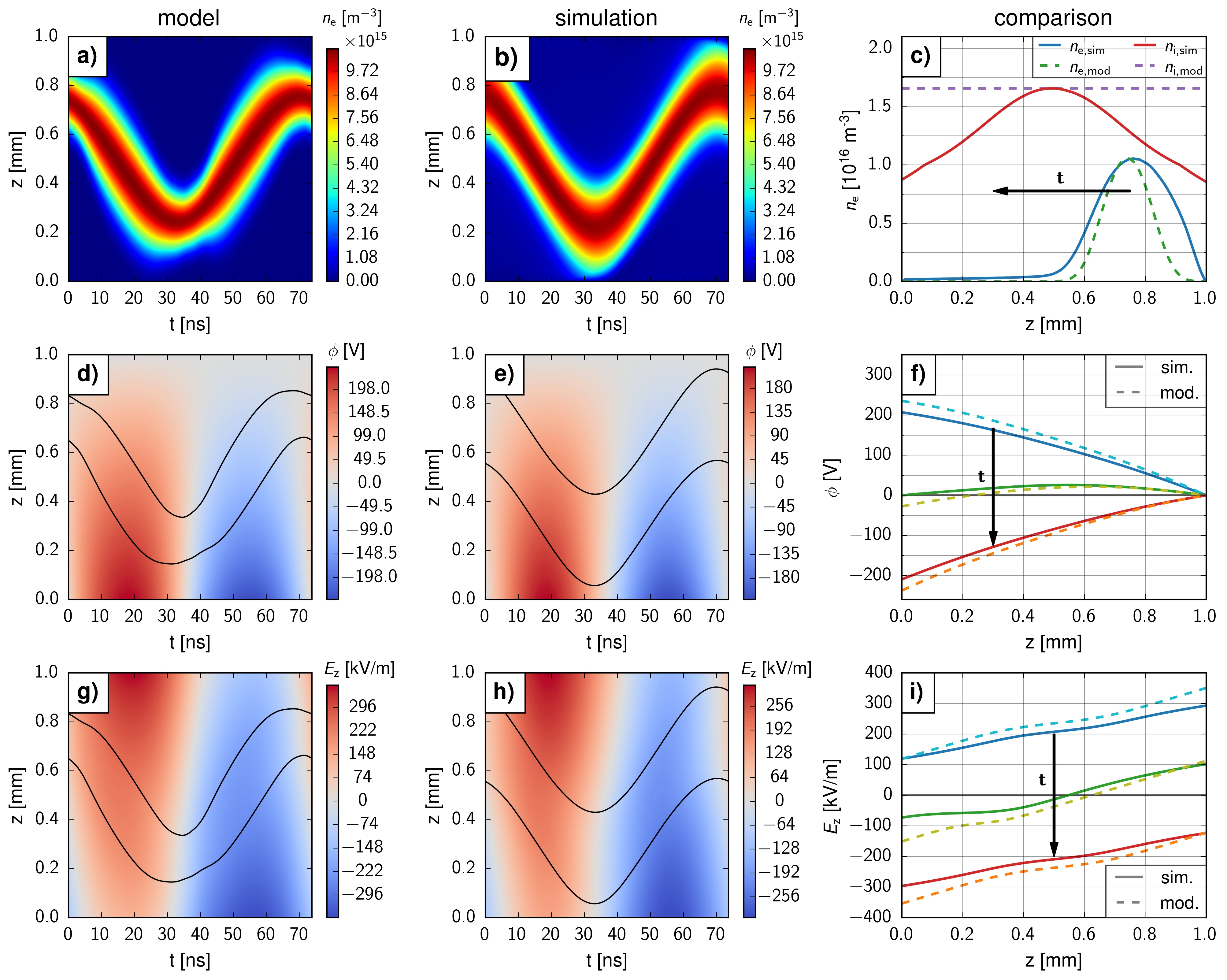}
		\caption{Comparison of the analytical model in physical units and laboratory coordinates with the results of the hybrid PIC/MCC simulation.
			The left column shows for the analytical model~the temporally and spatially resolved  a) electron density $n_\mathrm{e}$, d)  electric potential $\phi$, g) the electric field~$E_z$.
			The middle column shows the same quantities from the simulation.
			The right column compares~snapshots.
			Movie 4 provides an animation. 
			The black lines indicate the estimated position of the electron group.
			The data are obtained from the hybrid PIC/MCC simulation with the parameters: $V_\mathrm{RF} = 220\,$V, $x_\mathrm{N\textsubscript{2}\textsuperscript{+}} = 0.01\,$,  $\gamma_\mathrm{i,He\textsuperscript{+}} = \gamma_\mathrm{i,He\textsubscript{2}\textsuperscript{+}} = 0.25$, $\gamma_\mathrm{i,N\textsubscript{2}\textsuperscript{+}} = 0.1$.}
		\label{fig:phys_solutions}
	\end{figure}			
	
Figure \ref{fig:phys_solutions} compares the two models in terms of the temporally and spatially resolved electron density $n_\mathrm{e}$, electric potential $\phi$, and electric field $E_z$.
The left column presents spatially and temporally resolved results of the analytical model.
The electron density $n_\mathrm{e}$ is shown in panel a), the electric potential $\phi$ is shown in panel d), and the electric field $E_{z}$ is shown in panel g).
The mid column displays temporally and spatially resolved simulation results for the same quantities (b): $n_\mathrm{e}$, (e): $\phi$, (h): $E_{z}$.
The right column presents temporal snapshots of the comparison between the analytical model and the simulation results (c): densities, (f): $\phi$, (i): $E_z$.
Overall, an excellent qualitative and a satisfactory quantitative agreement is observed.
This agreement and details of the following discussion are best seen when looking at the animated version of figure \ref{fig:phys_solutions} given in the attached movie 4.
Deviations are most pronounced when the soliton-like structure approaches either of the surfaces.
All deviations are plausibly explained by the fact that the analytical model assumes a constant ion density and neglects the electron absorption at the  walls.
The assumption of a constant ion density leads to lesser accurate numerical values for the electric potential $\phi$ and field $E_z$.
The neglect of wall interactions causes the modeled soliton to get stretched at the walls (c.f., fig. \ref{fig:phys_solutions} a), b), and c)). \par

A detailed analysis of figure \ref{fig:phys_solutions} reveals that the difference of the model behaves as one would expect based on Poisson's equation \eqref{eq:Poisson}.
The electron density $n_\mathrm{e}$ and the electric field $E_z$ are connected by one integration, and the electric potential $\phi$ is calculated by integrating $n_\mathrm{e}$ twice.
Each integration by its mathematical nature results in a smoother quantity and differences between the integrands become increasingly irrelevant.
Accordingly, the electron density shows both qualitatively and quantitatively the strongest deviations from the simulation results (c.f., fig. \ref{fig:phys_solutions} c)), and the electric potential $\phi$ bears the best agreement (i.e., there is just a slight quantitative deviation - c.f., fig. \ref{fig:phys_solutions} i)).
Moreover, the qualitative agreement between the temporally and spatially resolved profiles of the electric potential $\phi$ and the field $E_z$ (compare fig. \ref{fig:phys_solutions} d) to e) and fig. \ref{fig:phys_solutions} g) to h)) makes the respective contours nearly indistinguishable.\par

In summary and supported by both simulation and theoretical analysis, the following important physical statements can be made:
\begin{itemize}
    \item There is indeed a non-neutral regime of RF-driven plasma jets.
    In the whole system, the phase-averaged electron density is considerably smaller than the ion density.
    A quasi-neutral bulk does not develop.
    Instead, the electrons organize themselves in the form of a soliton-like structure with a maximum that lies below the ion density.  
    This reflects that the electron soliton is not much larger than the Debye length, but scales with it.

    \item The observed soliton structure is stable.
    In the simulation, this is evident from the fast recovery of the form after distortions by the boundary.
    In the analytical model, this was formally proved by means of stability analysis.
    Physically, the soliton is (in the co-moving frame) an example of a Poisson-Boltzmann structure.
    
    \item Due to being related to the dynamics of an equilibrium point (i.e., the COM $\mathcal{Z}(t)$), the soliton dynamics govern the electron dynamics.
    The previously discussed fast recovery from distortions means that electrons rapidly close up to the soliton.

    \item One remarkable prediction of the simulation is the weak spatial gradient of the electron temperature. The assumption of a spatially constant $T_\mathrm{e}$ is thus well justified.
    The analytical model assumes that $T_\mathrm{e}$ it is also temporally constant; this is clearly an oversimplification that deserves attention. 
\end{itemize}

In conclusion, the physical statements extracted from the model answer the previously raised key questions: (i) How does the soliton-like structure form?, and (ii) Why do the electrons behave in a uniform fashion?

 \pagebreak


\section{Summary and outlook} \label{summary}
	
The purpose of this study was to describe and analyze  the non-neutral regime of capacitively coupled atmospheric pressure plasma jets via the combined use of numerical simulation and analytical modeling. In the simulation section, a self-developed hybrid PIC/MCC code was employed.
It was found that the discharge does not develop a stationary quasi-neutral bulk. Instead, the electrons form a mobile Gaussian-shaped structure that we termed a soliton. A characteristic feature of this soliton is its ability to recover its form when perturbed by an interaction with the walls. 
Another peculiar observation was that the electron temperature within the soliton is temporally modulated but spatially constant.
	
For a better insight into the underlying dynamics, we analyzed a simplified and thus transparent model of the electron dynamics. In a current-free (i.e., co-moving)~frame, 
the equilibrium solutions (solitons) were found to follow a Poisson-Boltzmann equation. \linebreak
Employing Lyapunov's second argument, we could show that the
solitons are stable. 
Linear stability analysis allowed to investigate the dynamic behavior of perturbations of the equilibrium solution. We found that all distortions decay on timescales comparable to the dielectric relaxation time,
way below the RF period. The Boltzmann equilibrium character of the 
solitons explains also the observed uniform heating.

The work closed with a comparison between the results of the hybrid PIC/MCC simulation and a parameter fitted solution of the analytical description.
Despite its simplified nature, the~analytical model performed quite well.
Deviations between the models could be directly traced back to the simplifying assumptions of the analytical model.
In the near future, we plan to expand our work on the non-neutral discharge regime. We plan to establish precise criteria for the non-neutral regime,
to distinguish it from the classical discharge modes \cite{Shaper2009, Kelly2014, Korolov2019, Korolov2020}. 
During this mode transition, close attention will have to be paid to classical concepts such as the boundary sheath and quasi-neutrality.
Both have no meaning within the non-neutral regime, and preliminary evidence suggests that their definition will be tricky in the transition regime when $n_0$ approaches unity.
Also the other limit poses open questions: The analytical model
yields solutions for arbitrarily small parameters $n_0$.
Is~there a lower limit to the value of $n_0$ below which stable 
discharge solutions cannot exist?\par

As a final remark, we stress that we would be interested in experimental validation of our theoretical speculation. To the best of our knowledge,
there are currently no published experimental studies that confirm the existence of
a non-neutral discharge regime of RF driven atmospheric pressure plasma jets.
	
\section*{Acknowledgement} Funding by the German Research Foundation DFG,  project 327886311 (CRC 1316), and the National Research, Development and Innovation Office (Hungary), project K134462, is gratefully acknowledged.

\clearpage


\end{document}